\documentclass[tightenlines, aps, prd, preprint, draft, showpacs]{revtex4}

\begin{document}

\title{Regularization parameters for the self-force of a scalar particle
in a general orbit about a Schwarzschild black hole}

\author{Dong-Hoon Kim}

\affiliation{Department of Physics, PO Box 118440, University of Florida,
Gainesville, FL 32611-8440}

\date{\today{}}

\begin{abstract}
The interaction of a charged particle with its own field results in
the \emph{self-force} on the particle, which includes but is more
general than the radiation reaction force. In the vicinity of the
particle in curved spacetime, one may follow Dirac and split the retarded
field of the particle into two parts, (1) the singular source field,
$\sim q/r$, and (2) the regular remainder field. The singular source
field exerts no force on the particle, and the self-force is entirely
caused by the regular remainder. We describe an elementary multipole
decomposition of the singular source field which is an important step
in the calculation of the self-force on a scalar-charged particle
orbiting a Schwarzschild black hole.
\end{abstract}
\maketitle

\section{INTRODUCTION\label{sec:INTRODUCTION}}

According to the equivalence principle in general relativity, a particle
of infinitesimal mass orbits a black hole of large mass along a geodesic
worldline $\Gamma$ in the background spacetime determined by the
large mass alone. For a particle of small but finite mass, the orbit
is no longer a geodesic in the background of the large mass because
the particle perturbs the spacetime geometry. This perturbation due
to the presence of the smaller mass modifies the orbit of the particle
from an original geodesic in the background. The difference of the
actual orbit from a geodesic in the background is said to result from
the interaction of the moving particle with its own gravitational
field, which is called a \emph{self-force} \cite{detweiler-whiting(03)}.

Historically, Dirac \cite{dirac(38)} first gave the analysis of the
self-force for the electromagnetic field of a particle in flat spacetime.
He was able to approach the problem in a perturbative scheme by allowing
the particle's size to remain finite and invoking the conservation
of the stress-energy tensor inside a narrow world tube surrounding
the particle's worldline. Dewitt and Brehme \cite{dewitt-brehme(60)}
extended Dirac's problem to curved spacetime. Mino, Sasaki, and Tanaka
\cite{mino-sasaki-tanaka(97)} generalized it for the gravitational
field self-force. Quinn and Wald \cite{quinn-wald(97)} and Quinn
\cite{quinn(00)} worked out similar schemes for the gravitational,
electromagnetic, and scalar field self-forces by taking an axiomatic
approach.

In Dirac's \cite{dirac(38)} flat spacetime problem, the retarded
field is decomposed into two parts: (\emph{i}) The first part is the
{}``mean of the advanced and retarded fields'' which is a solution
of the inhomogeneous field equation resembling the Coulomb $q/r$
piece of the scalar potential near the particle. (\emph{ii}) The second
part is a {}``radiation'' field which is a homogeneous solution
of Maxwell's equations. Dirac describes the self-force as the interaction
of the particle with the radiation field, a well-defined solution
of the vacuum field equations.

In the analyses of the self-force in curved spacetime, the Hadamard
form of Green's function \cite{dewitt-brehme(60)} is employed to
describe the retarded field of the particle. Traditionally, taking
the scalar field case for example, the retarded Green's function $G^{\mathrm{ret}}(p,p')$
is divided into \emph{direct} and \emph{tail} parts: (\emph{i}) The
direct part has support only on the past null cone of the field point
$p$. (\emph{ii}) The tail part has support inside the past null cone
due to the presence of the curvature of spacetime. Accordingly, the
self-force on the particle consists of two pieces: (\emph{i}) The
first piece comes from the direct part of the field and the acceleration
of the worldline in the background geometry; this corresponds to Abraham-Lorentz-Dirac
(ALD) force in flat spacetime. (\emph{ii}) The second piece comes
from the tail part of the field and is present in curved spacetime.
Thus, the description of the self-force in curved spacetime reduces
to Dirac's result in the flat spacetime limit. In this approach, the
self-force is considered to result via \begin{equation}
\mathcal{F}_{a}=q\nabla_{a}\psi,\label{eq:1}\end{equation}
 from the interaction of the particle with the quantity \cite{detweiler-whiting(03)}\begin{eqnarray}
\psi^{\mathrm{self}} & = & \psi^{\mathrm{ret}}-\psi^{\mathrm{dir}}\nonumber \\
 & = & -\left[\frac{{qu(p,p')}}{2\dot{{\sigma}}}\right]_{\tau_{\mathrm{ret}}}^{\tau_{\mathrm{adv}}}-q\int_{-\infty}^{\tau_{\mathrm{ret}}}v\left[p,p'(\tau')\right]d\tau'.\label{psi-self}\end{eqnarray}
 The quantities $u$, $v$ and $\sigma$ are familiar from the Hadamard
expansion of a Green's function \cite{dewitt-brehme(60)}.

Although this traditional approach provides adequate methods to compute
the self-force, it does not share the physical simplicity of Dirac's
analysis where the force is described entirely in terms of an identifiable,
vacuum solution of the field equations: unlike Dirac's radiation field,
the $\psi^{\mathrm{self}}$ in Eq.~(\ref{psi-self}) is not a solution
of the vacuum field equation $\nabla^{2}\psi=0$. Moreover, the integral
term in $\psi^{\mathrm{self}}$ comes from the tail part of the Green's
function and is generally not differentiable on the worldline if the
Ricci scalar of the background is not zero (similarly, the electromagnetic
potential $A_{a}^{\mathrm{tail}}$ and the gravitational metric perturbation
$h_{ab}^{\mathrm{tail}}$ are not differentiable at the point of the
particle unless $\left(R_{ab}-\frac{{1}}{6}g_{ab}R\right)u^{b}$ and
$R_{cadb}u^{c}u^{d}$, respectively are zero in the background \cite{detweiler-m-w(03)}).
Thus, some version of averaging process must be invoked to make sense
of the self-force.

In this paper we use an alternative method to split the retarded field
$\psi^{\mathrm{ret}}$ in curved spacetime which is similar to Dirac's
and suggested by Ref.~\cite{detweiler-whiting(03)}: (\emph{i}) The
\emph{singular source field} $\psi^{\mathrm{S}}$ is an inhomogeneous
field similar to the tidally distorted Coulomb field and exerts no
force on the particle. (\emph{ii}) The \emph{regular remainder field}
$\psi^{\mathrm{R}}$ is a homogeneous solution of the field equation,
analogous to Dirac's radiation field, and is entirely responsible
for the self-force. This alternative split is reviewed briefly in
Section~\ref{sec:DECOMPOSITION-OF-THE}.

In Section~\ref{sec:MODE-SUM-DECOMPOSITION-AND} we give a brief
overview of the mode-sum decomposition scheme to evaluate the self-force
\cite{barack-ori(00)}. We consider a particle with a scalar charge
$q$ in general motion about a Schwarzschild black hole. A spherical
harmonic decomposition provides the multipole components of of both
$\psi^{\mathrm{ret}}$ and $\psi^{\mathrm{S}}$. Then, the mode by
mode sum of the difference of these components determines $\psi^{\mathrm{R}}$
and, thence, the self-force. The multipole components of $\psi^{\mathrm{ret}}$
can be determined numerically while the multipole components of $\psi^{\mathrm{S}}$
are derived analytically. In particular, the multipole moments of
$\psi^{\mathrm{S}}$ are generically referred to as \emph{regularization
parameters} \cite{barack-ori(00)}. This paper focuses on the analytical
task of finding these regularization parameters. Our analytical results
are summarized at the end of the Section in Eqs.~(\ref{eq:10})-(\ref{eq:17}).
These results are in agreement with those of Barack, Ori, Mino, Nakano,
and Sasaki \cite{barack-ori(02),mino-nakano-sasaki(02),barack-mino-n-o-s(02)}.

The description of $\psi^{\mathrm{S}}$ becomes particularly simple
in a specially chosen co-moving frame: the THZ normal coordinates,
introduced by Thorne and Hartle \cite{thorne-hartle(85)} and extended
by Zhang \cite{zhang(86)}, are locally inertial on a geodesic. In
Section~\ref{sec:DETERMINATION-OF-psiS} we obtain a simple form
for $\psi^{\mathrm{S}}$ using the THZ coordinates and then re-express
it in terms of the background Schwarzschild coordinates.

Section~\ref{sec:-REGULARIZATION-PARAMETERS} outlines our derivation
of the regularization parameters which, while not elementary, appears
to us to be more compact than the derivations of others \cite{barack-ori(02),mino-nakano-sasaki(02)}.

Appendix \ref{hyper} provides some mathematical details concerning
the hypergeometric functions and the different representations of
the regularization parameters in connection with them.

\textbf{Notation:} $\left(t,\, r,\,\theta,\,\phi\right)$ are the
usual Schwarzschild coordinates. The particle moves along a worldline
$\Gamma$, parameterized by the proper time $\tau$, The points $p$
and $p'$ refer to a field point and a source point, respectively,
on the worldline $\Gamma$ of the particle. In the coincidence limit
$p\rightarrow p'$. The coordinates $\left(T,\, X,\, Y,\, Z\right)$
are intermediate coordinates derived from the Schwarzschild coordinates,
while $\left(\mathcal{T},\,\mathcal{X},\,\mathcal{Y},\,\mathcal{Z}\right)$
are the THZ coordinates attached to the worldline $\Gamma$ of the
particle with $\rho\equiv\sqrt{{\mathcal{X}^{2}+\mathcal{Y}^{2}+\mathcal{Z}^{2}}}$.

\section{DECOMPOSITION OF THE RETARDED FIELD\label{sec:DECOMPOSITION-OF-THE}}

The recent analysis of the Green's function decomposition by Detweiler
and Whiting \cite{detweiler-whiting(03)} shows a method to split
the retarded field into two parts \begin{equation}
\psi^{\mathrm{ret}}=\psi^{\mathrm{S}}+\psi^{\mathrm{R}},\label{eq:2}\end{equation}
 where $\psi^{\mathrm{S}}$ and $\psi^{\mathrm{R}}$ are the singular
source field and the regular remainder field, respectively. The source
function for a point particle on the worldline $\Gamma$ is \begin{equation}
\varrho(p)=q\int(-g)^{-1/2}\delta^{4}(p-p'(\tau'))d\tau'.\end{equation}
 The singular source field $\psi^{\mathrm{S}}$ is an inhomogeneous
solution of the scalar field equation \begin{equation}
\nabla^{2}\psi=-4\pi\varrho\label{eq:3}\end{equation}
 in the neighborhood of the particle, just as $\psi^{\mathrm{ret}}$
is. And $\psi^{\mathrm{S}}$ is determined in the neighborhood of
the particle's worldline entirely by local analysis. $\psi^{\mathrm{R}}$,
defined by Eq.~(\ref{eq:2}), is then necessarily a homogeneous solution
and is therefore expected to be differentiable on $\Gamma$. According
to Ref.~\cite{detweiler-whiting(03)}, $\psi^{\mathrm{R}}$ will
formally give the correct self-force when substituted on the right
hand side of Eq.~(\ref{eq:1}) in place of $\psi^{\mathrm{self}}$.
In this paper we adopt this decomposition, and determine an analytical
approximation, via a multipole expansion, of $\psi^{\mathrm{S}}$,
which is to be subtracted from $\psi^{\mathrm{ret}}$ for an explicit
computation of the self-force.

\section{MODE-SUM DECOMPOSITION AND REGULARIZATION PARAMETERS \label{sec:MODE-SUM-DECOMPOSITION-AND}}

By Eq.~(\ref{eq:1}) the self-force can be formally evaluated from
\begin{eqnarray}
\mathcal{F}_{a}^{\mathrm{self}} & = & \lim_{p\rightarrow p'}\left[\mathcal{F}_{a}^{\mathrm{ret}}(p)-\mathcal{F}_{a}^{\mathrm{S}}(p)\right]=\mathcal{F}_{a}^{\mathrm{R}}(p')\nonumber \\
 & = & q\lim_{p\rightarrow p'}\nabla_{a}(\psi^{\mathrm{ret}}-\psi^{\mathrm{S}})=q\nabla_{a}\psi^{\mathrm{R}},\label{eq:4}\end{eqnarray}
 where $p'$ is the event on $\Gamma$ where the self-force is to
be determined and $p$ is an event in the neighborhood of $p'$. For
use of this equation, both $\mathcal{F}_{a}^{\mathrm{ret}}(p)$ and
$\mathcal{F}_{a}^{\mathrm{S}}(p)$ would be expanded into multipole
$\ell$-modes, with $\mathcal{F}_{\ell a}^{\mathrm{ret}}(p)$ determined
numerically.

For the Schwarzschild spacetime, the source function $\varrho(p)$
is expanded in terms of spherical harmonics, and a similar expansion
for $\psi^{\mathrm{ret}}$ is \begin{equation}
\psi^{\mathrm{ret}}={\displaystyle \sum_{\ell m}}\psi_{\ell m}^{\mathrm{ret}}(r,t)Y_{\ell m}(\theta,\phi),\label{eq:5}\end{equation}
 where $\psi_{\ell m}^{\mathrm{ret}}(r,t)$ is found numerically.
The individual components $\psi_{\ell m}^{\mathrm{ret}}$ in this
expansion are finite at the location of the particle even though their
sum is singular. Then, the $\ell$ component $\mathcal{F}_{\ell a}^{\mathrm{ret}}$
is finite \begin{equation}
\mathcal{F}_{\ell a}^{\mathrm{ret}}=q\nabla_{a}\sum_{m}\psi_{\ell m}^{\mathrm{ret}}Y_{\ell m}.\label{eq:6}\end{equation}
 The singular source field $\psi^{\mathrm{S}}$ is determined analytically
in the neighborhood of the particle's worldline via local analysis
(see Section \ref{sec:DETERMINATION-OF-psiS}) and its mode-sum decomposition
provides \begin{equation}
\mathcal{F}_{\ell a}^{\mathrm{S}}=q\nabla_{a}\sum_{m}\psi_{\ell m}^{\mathrm{S}}Y_{\ell m},\label{eq:7}\end{equation}
 which is also finite at the location of the particle. Eqs. (\ref{eq:4}),
(\ref{eq:6}) and (\ref{eq:7}) now imply that \begin{eqnarray}
\mathcal{F}_{a}^{\mathrm{self}} & = & \sum_{\ell}\lim_{p\rightarrow p'}\left[\mathcal{F}_{\ell a}^{\mathrm{ret}}(p)-\mathcal{F}_{\ell a}^{\mathrm{S}}(p)\right]\nonumber \\
 & = & q\sum_{\ell}\lim_{p\rightarrow p'}\nabla_{a}\sum_{m}(\psi_{\ell m}^{\mathrm{ret}}-\psi_{\ell m}^{\mathrm{S}})Y_{\ell m}\label{eq:8}\end{eqnarray}
 evaluated at the location of the particle.

We follow Barack and Ori \cite{barack-ori(00)} in defining the regularization
parameters, except that the singular source field $\psi^{\mathrm{S}}$
is used in place of $\psi^{\mathrm{dir}}$, \begin{equation}
\lim_{p\rightarrow p'}\mathcal{F}_{\ell a}^{\mathrm{S}}=\left(\ell+\frac{{1}}{2}\right)A_{a}+B_{a}+\frac{{C_{a}}}{\ell+\frac{{1}}{2}}+O(\ell^{-2}).\label{eq:9}\end{equation}
 In Section \ref{sec:-REGULARIZATION-PARAMETERS} these regularization
parameters are derived from the multipole components of $\nabla_{a}\psi^{\mathrm{S}}$
evaluated at the source point, \begin{equation}
A_{t}=\mathrm{sgn}(\Delta)\frac{{q^{2}}}{r_{\mathrm{o}}^{2}}\frac{{\dot{r}}}{1+J^{2}/r_{\mathrm{o}}^{2}},\label{eq:10}\end{equation}
\begin{equation}
A_{r}=-\mathrm{sgn}(\Delta)\frac{{q^{2}}}{r_{\mathrm{o}}^{2}}\frac{{E}\left(1-\frac{{2M}}{r_{\mathrm{o}}}\right)^{-1}}{1+J^{2}/r_{\mathrm{o}}^{2}},\label{eq:11}\end{equation}
\begin{equation}
A_{\phi}=0,\label{eq:12}\end{equation}
\begin{equation}
B_{t}=\frac{{q^{2}}}{r_{\mathrm{o}}^{2}}E\dot{r}\left[\frac{{F_{3/2}}}{\left(1+J^{2}/r_{\mathrm{o}}^{2}\right)^{3/2}}-\frac{{3F_{5/2}}}{2\left(1+J^{2}/r_{\mathrm{o}}^{2}\right)^{5/2}}\right],\label{eq:13}\end{equation}
\begin{equation}
B_{r}=\frac{{q^{2}}}{r_{\mathrm{o}}^{2}}\left\{ -\frac{{F_{1/2}}}{\left(1+J^{2}/r_{\mathrm{o}}^{2}\right)^{1/2}}+\frac{{\left[1-2\left(1-\frac{{2M}}{r_{\mathrm{o}}}\right)^{-1}\dot{{r}}^{2}\right]{F_{3/2}}}}{2\left(1+J^{2}/r_{\mathrm{o}}^{2}\right)^{3/2}}+\frac{{3}\left(1-\frac{{2M}}{r_{\mathrm{o}}}\right)^{-1}\dot{{r}}^{2}F_{5/2}}{2\left(1+J^{2}/r_{\mathrm{o}}^{2}\right)^{5/2}}\right\} ,\label{eq:14}\end{equation}
\begin{equation}
B_{\phi}=\frac{{q^{2}}}{J}\dot{r}\left[\frac{{F_{1/2}-F_{3/2}}}{\left(1+J^{2}/r_{\mathrm{o}}^{2}\right)^{1/2}}+\frac{{3(F_{5/2}-F_{3/2})}}{2\left(1+J^{2}/r_{\mathrm{o}}^{2}\right)^{3/2}}\right],\label{eq:15}\end{equation}
\begin{equation}
C_{t}=C_{r}=C_{\phi}=0,\label{eq:16}\end{equation}
\begin{equation}
A_{\theta}=B_{\theta}=C_{\theta}=0,\label{eq:17}\end{equation}
 where $\Delta\equiv r-r_{\mathrm{o}}$, $E\equiv-u_{t}=(1-2M/r_{\mathrm{o}})\left(dt/d\tau\right)_{\mathrm{o}}$
($\tau$: proper time) and $J\equiv u_{\phi}=r_{\mathrm{o}}^{2}\left(d\phi/d\tau\right)_{\mathrm{o}}$
are the conserved energy and angular momentum, respectively, and $\dot{r}\equiv u^{r}=\left(dr/d\tau\right)_{\mathrm{o}}$.
The subscript ${}_{\mathrm{o}}$ denotes evaluation at the location
of the particle. Also, shorthand for the hypergeometric function is
$F_{p}\equiv{}_{2}F_{1}\left(p,\frac{{1}}{2};1;J^{2}/(r_{\mathrm{o}}^{2}+J^{2})\right)$
(see Appendix \ref{hyper} for more details about the hypergeometric
functions and the representations of the regularization parameters
in terms of them).

\section{DETERMINATION OF $\psi^{\mathrm{S}}$ in locally inertial coordinates
\label{sec:DETERMINATION-OF-psiS}}

In the vicinity of an event $p'$ on a timelike worldline $\Gamma$,
physics is most easily described in terms of locally inertial coordinates,
where the time coordinate $\mathcal{T}$ on $\Gamma$ is equal to
the the proper time, and the orthogonal Cartesian-like spatial coordinates
are $\left(\mathcal{X},\,\mathcal{Y},\,\mathcal{Z}\right)$ and centered
on $\Gamma$. At $p'$, with locally inertial coordinates the spacetime
metric is just the flat Minkowski metric, and all of the Christoffel
symbols vanish. Locally inertial coordinates are not unique and have
an ambiguity at $O(\rho^{3})$, where $\rho^{2}=\sqrt{\mathcal{X}^{2}+\mathcal{Y}^{2}+\mathcal{Z}^{2}}$.
For example, differences of $O(\rho^{3})$ distinguish Riemann normal
from Fermi normal coordinates \cite{MTW(73)}. For our purposes a
locally inertial coordinate system introduced by Thorne and Hartle
\cite{thorne-hartle(85)} and later extended by Zhang \cite{zhang(86)}
is particularly advantageous. In these \emph{THZ} coordinates Detweiler,
Messaritaki, and Whiting \cite{detweiler-m-w(03)} (cited henceforth
as Paper I) show that the scalar wave equation takes a simple form
and also that \begin{equation}
\psi^{\mathrm{S}}=q/\rho+O(\rho^{2}/\mathcal{R}^{3}),\label{eq:18}\end{equation}
 where $\mathcal{R}$ represents a length scale of the background
geometry (the smallest of the radius of curvature, the scale of inhomogeneities
and time scale for changes in curvature along $\Gamma$). Approximation
(\ref{eq:18}) is accurate enough for self-force regularization because
\begin{equation}
\nabla_{a}\psi^{\mathrm{S}}=\nabla_{a}(q/\rho)+O(\rho/\mathcal{R}^{3}),\end{equation}
 and the $O(\rho/\mathcal{R}^{3})$ remainder vanishes at the particle.

For the derivation of the regularization parameters from the multipole
components of $\nabla_{a}\psi^{\mathrm{S}}$, requires that $\rho$
in Eq.~(\ref{eq:18}) be expressed in terms of the coordinates of
the background geometry. Thus, we look for the relationship between
the original Schwarzschild coordinates $\left(t,\, r,\,\theta,\,\phi\right)$
and the THZ coordinates $\left(\mathcal{T},\,\mathcal{X},\,\mathcal{Y},\,\mathcal{Z}\right)$
associated with an event $p'$ on $\Gamma$. However, in Section \ref{sec:-REGULARIZATION-PARAMETERS}
we surprisingly find that \emph{any} locally inertial coordinate system
is sufficient to determine the regularization parameters quoted in
Eqs.~(\ref{eq:10})-(\ref{eq:17}), which are actually independent
of the $O(\rho^{3})$ terms in the THZ coordinates. An elementary
discussion in Weinberg \cite{weinberg(72)} determines this coordinate
transformation through terms of $O(\rho^{2})$ in two steps:\\
\\
(\emph{i}) Find inertial coordinates $X^{A}$ in the neighborhood
of the event $p'$ on $\Gamma$ in terms of a Taylor expansion of
the Schwarzschild coordinates $x^{a}$ about $p'$, where the Schwarzschild
coordinates at $p'$ are $x_{\mathrm{o}}^{a}$ and the subscript ${}_{\mathrm{o}}$
denotes evaluation at $p'$. Weinberg's \cite{weinberg(72)} Eq.~(3.2.12)
is \begin{equation}
X^{A}=X_{\mathrm{o}}^{A}+M^{A}{}_{a}(x^{a}-x_{\mathrm{o}}^{a})+\frac{{1}}{2}M^{A}{}_{a}\left.\Gamma_{bc}^{a}\right|_{\mathrm{o}}(x^{b}-x_{\mathrm{o}}^{b})(x^{c}-x_{\mathrm{o}}^{c})+O[(x-x_{\mathrm{o}})^{3}],\label{eq:19}\end{equation}
 where we may choose $X_{\mathrm{o}}^{A}=0$ and $M^{A}{}_{a}=\mathrm{diag}\left[M^{T}{}_{t},\, M^{X}{}_{r},\, M^{Y}{}_{\phi},\, M^{Z}{}_{\theta}\right]$
for convenience as this choice recenters and rescales the Schwarzschild
coordinates to $T=M^{T}{}_{t}(t-t_{\mathrm{o}})$, $X=M^{X}{}_{r}(r-r_{\mathrm{o}})$,
$Y=M^{Y}{}_{\phi}(\phi-\phi_{\mathrm{o}})$, $Z=M^{Z}{}_{\theta}(\theta-\theta_{\mathrm{o}})$.\\
\\
(\emph{ii}) Boost $X^{A}$ with $u^{A}$, the particle's four-velocity
at $p'$ as measured in this Cartesian frame, to obtain the final
coordinates $\mathcal{X}^{A'}$; \begin{eqnarray}
\mathcal{X}^{A'} & = & \Lambda^{A'}{}_{A}X^{A}\nonumber \\
 & = & \Lambda^{A'}{}_{A}\left[M^{A}{}_{a}(x^{a}-x_{\mathrm{o}}^{a})+\frac{{1}}{2}M^{A}{}_{a}\left.\Gamma_{bc}^{a}\right|_{\mathrm{o}}(x^{b}-x_{\mathrm{o}}^{b})(x^{c}-x_{\mathrm{o}}^{c})\right]+O[(x-x_{\mathrm{o}})^{3}],\label{eq:20}\end{eqnarray}
 where \begin{equation}
\Lambda^{A'}{}_{A}=\left[\begin{array}{cccc}
u^{T} & -u^{X} & -u^{Y} & -u^{Z}\\
 & 1+(u^{T}-1)(u^{X})^{2}/u^{2} & (u^{T}-1)u^{X}u^{Y}/u^{2} & (u^{T}-1)u^{X}u^{Z}/u^{2}\\
 &  & 1+(u^{T}-1)(u^{Y})^{2}/u^{2} & (u^{T}-1)u^{Y}u^{Z}/u^{2}\\
 & \textrm{sym } &  & 1+(u^{T}-1)(u^{Z})^{2}/u^{2}\end{array}\right]\label{boost}\end{equation}
 is the upper half of the symmetric matrix $\Lambda^{A'}{}_{A}$ with
$u^{2}\equiv(u^{X})^{2}+(u^{Y})^{2}+(u^{Z})^{2}$ \cite{jackson(99)}.\\

With the choice of $M^{T}{}_{t}=\left(1-2M/r_{\mathrm{o}}\right)^{1/2}$,
$M^{X}{}_{r}=\left(1-2M/r_{\mathrm{o}}\right)^{-1/2}$, $M^{Y}{}_{\phi}=r_{\mathrm{o}}\sin\theta_{\mathrm{o}}$
and $M^{Z}{}_{\theta}=-r_{\mathrm{o}}$, it follows that \begin{eqnarray}
g^{A'B'} & = & g^{ab}\frac{{\partial\mathcal{X}^{A'}}}{\partial x^{a}}\frac{{\partial\mathcal{X}^{B'}}}{\partial x^{b}}\nonumber \\
 & = & \eta^{A'B'}+O[(x-x_{\mathrm{o}})^{2}],\; x^{a}\rightarrow x_{\mathrm{o}}^{a},\label{eq:21}\end{eqnarray}
 so that\begin{equation}
\frac{{\partial g^{A'B'}}}{\partial\mathcal{X}^{C'}}=O[(x-x_{\mathrm{o}})],\; x^{a}\rightarrow x_{\mathrm{o}}^{a},\label{eq:22}\end{equation}
 Eqs.~(\ref{eq:21}) and (\ref{eq:22}) are the desired locally inertial
features for a particle in the Schwarzschild geometry at event $x_{\mathrm{o}}^{a}$
with four-velocity $u^{a}$.

To simplify the calculations, we confine the particle's orbit to the
equatorial plane $\theta_{\mathrm{o}}=\pi/2$ and have \begin{equation}
M^{A}{}_{a}=\mathrm{diag}\left[f^{1/2},\, f^{-1/2},\, r_{\mathrm{o}},\,-r_{\mathrm{o}}\right],\label{eq:23}\end{equation}
 where $f\equiv(1-\frac{{2M}}{r_{\mathrm{o}}})$. This constraint
to the equatorial plane makes $u^{Z}=0$ and we rewrite $u^{A}$\begin{equation}
u^{A}\equiv(u^{T},\, u^{X},\, u^{Y},\, u^{Z})=\left(f^{-1/2}E,\, f^{-1/2}\dot{{r}},\,\frac{{J}}{r_{\mathrm{o}}},\,0\right),\label{eq:24}\end{equation}
 in terms of the Schwarzschild coordinates and the constants of motion:
$E\equiv-u_{t}=f\left(dt/d\tau\right)_{\mathrm{o}}$ and $J\equiv u_{\phi}=r_{\mathrm{o}}^{2}\left(d\phi/d\tau\right)_{\mathrm{o}}$
are the conserved energy and angular momentum, respectively, and $\dot{r}\equiv u^{r}=\left(dr/d\tau\right)_{\mathrm{o}}$.
From this it follows that $u^{2}=f^{-1}E^{2}-1$ and we have \begin{equation}
\Lambda^{A'}{}_{A}=\left[\begin{array}{cccc}
f^{-1/2}E & -f^{-1/2}\dot{{r}} & -J/r_{\mathrm{o}} & 0\\
 & 1+\dot{{r}}^{2}/(f^{1/2}E+f) & J\dot{{r}}/[r_{\mathrm{o}}(E+f^{1/2})] & 0\\
 &  & 1+J^{2}/[r_{\mathrm{o}}^{2}(f^{-1/2}E+1)] & 0\\
 & \textrm{sym} &  & 1\end{array}\right].\label{eq:25}\end{equation}

Now we are able to express $\rho^{2}$ in terms of the Schwarzschild
coordinates using Eq.~(\ref{eq:20}) and obtain \begin{eqnarray}
\rho^{2}=\mathcal{X}^{I}\mathcal{X}_{I} & = & \delta_{IJ}\Lambda^{I}{}_{C}\Lambda^{J}{}_{D}M^{C}{}_{c}M^{D}{}_{d}\left[(x^{c}-x_{\mathrm{o}}^{c})(x^{d}-x_{\mathrm{o}}^{d})+\left.\Gamma_{ab}^{c}\right|_{\mathrm{o}}(x^{a}-x_{\mathrm{o}}^{a})(x^{b}-x_{\mathrm{o}}^{b})(x^{d}-x_{\mathrm{o}}^{d})\right]\nonumber \\
 &  & +O[(x-x_{\mathrm{o}})^{4}],\label{eq:26}\end{eqnarray}
 where $I,\, J=1,\,2,\,3$. Then, after a substitution from Eqs.~(\ref{eq:23})
and (\ref{eq:25}), $\rho^{2}$ becomes \begin{eqnarray}
\rho^{2} & = & (E^{2}-f)(t-t_{\mathrm{o}})^{2}-\frac{{2E\dot{{r}}}}{f}(t-t_{\mathrm{o}})(r-r_{\mathrm{o}})-2EJ(t-t_{\mathrm{o}})(\phi-\phi_{\mathrm{o}})\nonumber \\
 &  & +\frac{{1}}{f}\left(1+\frac{{\dot{{r}}^{2}}}{f}\right)(r-r_{\mathrm{o}})^{2}+\frac{{2J\dot{{r}}}}{f}(r-r_{\mathrm{o}})(\phi-\phi_{\mathrm{o}})+(r_{\mathrm{o}}^{2}+J^{2})(\phi-\phi_{\mathrm{o}})^{2}+r_{\mathrm{o}}^{2}\left(\theta-\frac{{\pi}}{2}\right)^{2}\nonumber \\
 &  & -\frac{{ME\dot{{r}}}}{r_{\mathrm{o}}^{2}}(t-t_{\mathrm{o}})^{3}+\frac{{M}}{r_{\mathrm{o}}^{2}}\left(-1+\frac{{2E^{2}}}{f}+\frac{{\dot{{r}}^{2}}}{f}\right)(t-t_{\mathrm{o}})^{2}(r-r_{\mathrm{o}})+\frac{{MJ\dot{{r}}}}{r_{\mathrm{o}}^{2}}(t-t_{\mathrm{o}})^{2}(\phi-\phi_{\mathrm{o}})\nonumber \\
 &  & -\frac{{ME\dot{{r}}}}{f^{2}r_{\mathrm{o}}^{2}}(t-t_{\mathrm{o}})(r-r_{\mathrm{o}})^{2}-\frac{{2(r_{\mathrm{o}}-M)EJ}}{fr_{\mathrm{o}}^{2}}(t-t_{\mathrm{o}})(r-r_{\mathrm{o}})(\phi-\phi_{\mathrm{o}})\nonumber \\
 &  & +r_{\mathrm{o}}E\dot{{r}}(t-t_{\mathrm{o}})(\phi-\phi_{\mathrm{o}})^{2}+r_{\mathrm{o}}E\dot{{r}}(t-t_{\mathrm{o}})\left(\theta-\frac{{\pi}}{2}\right)^{2}\nonumber \\
 &  & -\frac{{M}}{f^{2}r_{\mathrm{o}}^{2}}\left(1+\frac{{\dot{{r}}^{2}}}{f}\right)(r-r_{\mathrm{o}})^{3}+\frac{{(2r_{\mathrm{o}}-5M)J\dot{{r}}}}{f^{2}r_{\mathrm{o}}^{2}}(r-r_{\mathrm{o}})^{2}(\phi-\phi_{\mathrm{o}})\nonumber \\
 &  & +r_{\mathrm{o}}\left(1-\frac{{\dot{{r}}^{2}}}{f}+\frac{{2J^{2}}}{r_{\mathrm{o}}^{2}}\right)(r-r_{\mathrm{o}})(\phi-\phi_{\mathrm{o}})^{2}+r_{\mathrm{o}}\left(1-\frac{{\dot{{r}}^{2}}}{f}\right)(r-r_{\mathrm{o}})\left(\theta-\frac{{\pi}}{2}\right)^{2}\nonumber \\
 &  & -r_{\mathrm{o}}J\dot{{r}}(\phi-\phi_{\mathrm{o}})^{3}-r_{\mathrm{o}}J\dot{{r}}(\phi-\phi_{\mathrm{o}})\left(\theta-\frac{{\pi}}{2}\right)^{2}+O[(x-x_{\mathrm{o}})^{4}].\label{eq:27}\end{eqnarray}
 The substitution of Eq.~(\ref{eq:27}) into Eq.~(\ref{eq:18})
approximates $\psi^{\mathrm{S}}$ in terms of the Schwarzschild coordinates
and leads to the derivation of the regularization parameters in the
next section.

In the above analysis the $O[(x-x_{\mathrm{o}})^{3}]$ term in Eq.~(\ref{eq:20})
contributes to the $O[(x-x_{\mathrm{o}})^{4}]$ terms of $\rho^{2}$
in Eqs.~(\ref{eq:26}) and (\ref{eq:27}). To the level of accuracy
we desire for the mode-sum regularization parameters in this paper,
that is to say, to the determination of $C_{a}$-terms, it is not
necessary to specify the $O[(x-x_{\mathrm{o}})^{4}]$ term in $\rho^{2}$
and, hence, not necessary to specify the $O[(x-x_{\mathrm{o}})^{3}]$
terms in the spatial THZ coordinates $\mathcal{X}$, $\mathcal{Y}$,
$\mathcal{Z}$.

\section{REGULARIZATION PARAMETERS FOR A GENERAL ORBIT OF THE SCHWARZSCHILD
GEOMETRY\label{sec:-REGULARIZATION-PARAMETERS}}

In Section \ref{sec:DETERMINATION-OF-psiS}, we have seen that an
approximation to $\psi^{\mathrm{S}}$ is\begin{equation}
\psi^{\mathrm{S}}=q/\rho+O(\rho^{2}/\mathcal{R}^{3}).\label{eq:28}\end{equation}
 Following Paper I \cite{detweiler-m-w(03)}, the regularization parameters
can be determined from evaluating the multipole components of $\partial_{a}(q/\rho)$
($a=t,\, r,\,\theta,\,\phi$ for the Schwarzschild background). The
error, $O(\rho^{2}/\mathcal{R}^{3})$ in the above approximation is
disregarded since it gives no contribution to $\nabla_{a}\psi^{\mathrm{S}}$
as we take the {}``coincidence limit'', $x^{a}\rightarrow x_{\mathrm{o}}^{a}$
, where $x^{a}$ denotes a point in the vicinity of the particle and
$x_{\mathrm{o}}^{a}$ the location of the particle in the Schwarzschild
geometry.

In evaluating the multipole components of $\partial_{a}(q/\rho)$,
singularities are expected with certain terms. To help identify those
singularities, we introduce an order parameter $\epsilon$ which is
to be set to unity at the end of the calculation: we attach $\epsilon^{n}$
to each $O[(x-x_{\mathrm{o}})^{n}]$ part of $\rho^{2}$ in Eq.~(\ref{eq:27})
and re-express $\rho^{2}$ as \begin{equation}
\rho^{2}=\epsilon^{2}\mathcal{P}_{\mathrm{II}}+\epsilon^{3}\mathcal{P}_{\mathrm{III}}+\epsilon^{4}\mathcal{P}_{\mathrm{IV}}+O(\epsilon^{5}),\label{eq:29}\end{equation}
 where $\mathcal{P}_{\mathrm{II}}$, $\mathcal{P}_{\mathrm{III}}$,
and $\mathcal{P}_{\mathrm{IV}}$ represent the quadratic, cubic and
quartic order parts of $\rho^{2}$, respectively. Here we pretend
that the quartic part $\mathcal{P}_{\mathrm{IV}}$ is also specified:
this will help us to perform the structure analysis for $C_{a}$-terms
later in Subsection~\ref{sub:C-terms} when we prove that these regularization
parameters always vanish.

We express $\partial_{a}\left(1/\rho\right)$ in a Laurent series
expansion where every denominator of this expansion takes the form
of $\mathcal{P}_{\mathrm{II}}^{n/2}$ ($n=3,\,5,\,7$). Thus, $\mathcal{P}_{\mathrm{II}}$
plays an important role in the multipole decomposition, and the quadratic
part $\mathcal{P}_{\mathrm{II}}$, directly taken from Eq.~(\ref{eq:27}),
is not yet fully ready for this task. First, $\phi-\phi_{\mathrm{o}}$
must be decoupled from $r-r_{\mathrm{o}}$ so that each appears only
as an independent complete square. Coupling between $t-t_{\mathrm{o}}$
and $\phi-\phi_{\mathrm{o}}$ does not create difficulty in the decomposition.
Thus, we reshape the quadratic term of Eq.~(\ref{eq:27}) into \begin{eqnarray}
\mathcal{P}_{\mathrm{II}} & = & (E^{2}-f)(t-t_{\mathrm{o}})^{2}-\frac{{2E\dot{{r}}r_{\mathrm{o}}^{2}}}{f\left(r_{\mathrm{o}}^{2}+J^{2}\right)}(t-t_{\mathrm{o}})\Delta-2EJ(t-t_{\mathrm{o}})(\phi-\phi')\nonumber \\
 &  & +\frac{{E^{2}r_{\mathrm{o}}^{2}}}{f^{2}\left(r_{\mathrm{o}}^{2}+J^{2}\right)}\Delta^{2}+\left(r_{\mathrm{o}}^{2}+J^{2}\right)(\phi-\phi')^{2}+r_{\mathrm{o}}^{2}\left(\theta-\frac{{\pi}}{2}\right)^{2}\label{eq:33}\end{eqnarray}
 with \begin{equation}
\phi'\equiv\phi_{\mathrm{o}}-\frac{{J\dot{{r}}\Delta}}{f\left(r_{\mathrm{o}}^{2}+J^{2}\right)},\label{eq:34}\end{equation}
 where $\Delta\equiv r-r_{\mathrm{o}}$, and an identity $\dot{{r}}^{2}=E^{2}-f\left(1+J^{2}/r_{\mathrm{o}}^{2}\right)$
is used for simplifying the coefficient of $\Delta^{2}$. Here, taking
the coincidence limit $\Delta\rightarrow0$, we have $\phi'\rightarrow\phi_{\mathrm{o}}$.
This same idea is found in Mino, Nakano, and Sasaki~\cite{mino-nakano-sasaki(02)}.
Also, for the multipole decomposition the quadratic part must be analytic
and smooth over the entire two-sphere, and we write \begin{eqnarray}
\mathcal{P}_{\mathrm{II}} & = & (E^{2}-f)(t-t_{\mathrm{o}})^{2}-\frac{{2E\dot{{r}}r_{\mathrm{o}}^{2}}}{f\left(r_{\mathrm{o}}^{2}+J^{2}\right)}(t-t_{\mathrm{o}})\Delta-2EJ(t-t_{\mathrm{o}})\sin\theta\sin(\phi-\phi')\nonumber \\
 &  & +\frac{{E^{2}r_{\mathrm{o}}^{2}}}{f^{2}\left(r_{\mathrm{o}}^{2}+J^{2}\right)}\Delta^{2}+(r_{\mathrm{o}}^{2}+J^{2})\sin^{2}\theta\sin^{2}(\phi-\phi')+r_{\mathrm{o}}^{2}\cos^{2}\theta\nonumber \\
 &  & +O[(x-x_{\mathrm{o}})^{4}].\label{qdrt}\end{eqnarray}
 Here we have used the elementary approximations $\phi-\phi'=\sin(\phi-\phi')+O[(\phi-\phi')^{3}]$
and $1=\sin\theta+O[(\theta-\pi/2)^{2}]$.

To aid in the multipole decomposition we rotate the usual Schwarzschild
coordinates by following the approach of Barack and Ori~\cite{barack-ori(02)}
such that the coordinate location of the particle is moved from the
equatorial plane $\theta=\pi/2$ to the new polar axis. The new angles
$\Theta$ and $\Phi$ defined in terms of the usual Schwarzschild
angles are \begin{eqnarray}
\sin\theta\cos(\phi-\phi') & = & \cos\Theta\nonumber \\
\sin\theta\sin(\phi-\phi') & = & \sin\Theta\cos\Phi\nonumber \\
\cos\theta & = & \sin\Theta\sin\Phi.\label{eq:35}\end{eqnarray}
 Also, under this coordinate rotation, a spherical harmonic $Y_{\ell m}(\theta,\phi)$
becomes \begin{equation}
Y_{\ell m}(\theta,\phi)=\sum_{m'=-\ell}^{\ell}\alpha_{mm'}^{\ell}Y_{\ell m'}(\Theta,\Phi),\label{eq:36}\end{equation}
 where the coefficients $\alpha_{mm'}^{\ell}$ depend on the rotation
$(\theta,\phi)\rightarrow(\Theta,\Phi)$ as well as on $\ell$, $m$
and $m'$, and the index $\ell$ is preserved under the rotation \cite{mathews-walker(70)}.
As recognized in Ref.~\cite{barack-ori(02)}, there is a great advantage
of using the rotated angles $(\Theta,\Phi)$: after expanding $\partial_{a}(q/\rho)$
into a sum of spherical harmonic components, we take the coincidence
limit $\Delta\rightarrow0$, $\Theta\rightarrow0$. Then, finally
only the $m=0$ components contribute to the self-force at $\Theta=0$
because $Y_{\ell m}(0,\Phi)=0$ for $m\neq0$. Thus, the regularization
parameters of Eq.~(\ref{eq:9}) are just $(\ell,\, m=0)$ spherical
harmonic components of $\partial_{a}(q/\rho)$ evaluated at $x_{\mathrm{o}}^{a}$.

Now, with these rotated angles, $\mathcal{P}_{\mathrm{II}}$ is re-expressed
as \begin{eqnarray}
\mathcal{P}_{\mathrm{II}} & = & (E^{2}-f)(t-t_{\mathrm{o}})^{2}-\frac{{2E\dot{{r}}r_{\mathrm{o}}^{2}}}{f\left(r_{\mathrm{o}}^{2}+J^{2}\right)}(t-t_{\mathrm{o}})\Delta-2EJ(t-t_{\mathrm{o}})\sin\Theta\cos\Phi\nonumber \\
 &  & +2\left(r_{\mathrm{o}}^{2}+J^{2}\right)\left(1-\frac{{J^{2}\sin^{2}\Phi}}{r_{\mathrm{o}}^{2}+J^{2}}\right)\left[\frac{{r_{\mathrm{o}}^{2}E^{2}\Delta^{2}}}{2f^{2}\left(r_{\mathrm{o}}^{2}+J^{2}\right)^{2}\left(1-\frac{{J^{2}\sin^{2}\Phi}}{r_{\mathrm{o}}^{2}+J^{2}}\right)}+1-\cos\Theta\right]\nonumber \\
 &  & +O[(x-x_{\mathrm{o}})^{4}],\label{qdrt3}\end{eqnarray}
 where the elementary approximation $\sin^{2}\Theta=2(1-\cos\Theta)+O(\Theta^{4})$
is used. We may now define \begin{eqnarray}
\tilde{{\rho}}^{2} & \equiv & (E^{2}-f)(t-t_{\mathrm{o}})^{2}-\frac{{2E\dot{{r}}r_{\mathrm{o}}^{2}}}{f\left(r_{\mathrm{o}}^{2}+J^{2}\right)}(t-t_{\mathrm{o}})\Delta-2EJ(t-t_{\mathrm{o}})\sin\Theta\cos\Phi\nonumber \\
 &  & +2\left(r_{\mathrm{o}}^{2}+J^{2}\right)\left(1-\frac{{J^{2}\sin^{2}\Phi}}{r_{\mathrm{o}}^{2}+J^{2}}\right)\left[\frac{{r_{\mathrm{o}}^{2}E^{2}\Delta^{2}}}{2f^{2}\left(r_{\mathrm{o}}^{2}+J^{2}\right)^{2}\left(1-\frac{{J^{2}\sin^{2}\Phi}}{r_{\mathrm{o}}^{2}+J^{2}}\right)}+1-\cos\Theta\right].\label{A}\end{eqnarray}
 In particular, when fixing $t=t_{\mathrm{o}}$, we define \begin{equation}
\tilde{{\rho}}_{\mathrm{o}}^{2}\equiv\left.\tilde{{\rho}}^{2}\right|_{t=t_{\mathrm{o}}}=2\left(r_{\mathrm{o}}^{2}+J^{2}\right)\chi\left(\delta^{2}+1-\cos\Theta\right)\label{eq:37}\end{equation}
 with \begin{equation}
\chi\equiv1-\frac{{J^{2}\sin^{2}\Phi}}{r_{\mathrm{o}}^{2}+J^{2}}\label{eq:38}\end{equation}
 and \begin{equation}
\delta^{2}\equiv\frac{{r_{\mathrm{o}}^{2}E^{2}\Delta^{2}}}{2f^{2}\left(r_{\mathrm{o}}^{2}+J^{2}\right)^{2}\chi}.\label{eq:39}\end{equation}

Now we rewrite Eq.~(\ref{eq:29}) by replacing the original quadratic
part $\mathcal{P}_{\mathrm{II}}$ with $\tilde{{\rho}}^{2}$, \begin{equation}
\rho^{2}=\epsilon^{2}\tilde{{\rho}}^{2}+\epsilon^{3}\mathcal{P}_{\mathrm{III}}+\epsilon^{4}\mathcal{P}_{\mathrm{IV}}+O(\epsilon^{5}),\label{newrho2}\end{equation}
 where $\mathcal{P}_{\mathrm{IV}}$ now includes the additional quartic
order terms that have resulted from the replacement of $\mathcal{P}_{\mathrm{II}}$
by $\tilde{\rho}^{2}$. A Laurent series expansion of $\left.\partial_{a}(1/\rho)\right|_{t=t_{\mathrm{o}}}$
is \begin{equation}
\left.\partial_{a}\left(\frac{{1}}{\rho}\right)\right|_{t=t_{\mathrm{o}}}=-\frac{1}{2}\frac{{\left.\partial_{a}\left(\tilde{{\rho}}^{2}\right)\right|_{t=t_{\mathrm{o}}}}}{\tilde{{\rho}}_{\mathrm{o}}^{3}}\epsilon^{-2}+\left\{ -\frac{1}{2}\frac{{\left.\partial_{a}\mathcal{P}_{\mathrm{III}}\right|_{t=t_{\mathrm{o}}}}}{\tilde{{\rho}}_{\mathrm{o}}^{3}}+\frac{{3}}{4}\frac{{\left.\left[\partial_{a}\left(\tilde{{\rho}}^{2}\right)\right]\mathcal{P}_{\mathrm{III}}\right|_{t=t_{\mathrm{o}}}}}{\tilde{{\rho}}_{\mathrm{o}}^{5}}\right\} \epsilon^{-1}+O(\epsilon^{0}).\label{eq:30}\end{equation}

After the derivatives in Eq.~(\ref{eq:30}) are taken, the dependence
upon $\Theta$, $\Phi$ and $r$ may be removed in favor of $\tilde{{\rho}}_{\mathrm{o}}$,
$\chi$ and $\delta$ by use of Eqs.~(\ref{eq:37})--(\ref{eq:39}).
Then the three steps of (\emph{i}) a Legendre polynomial expansion
for the $\Theta$ dependence, while $r$ and $\Phi$ are held fixed,
followed by (\emph{ii}) an integration over $\Phi$, while $r$ is
held fixed, and finally (\emph{iii}) taking the limit $\delta\rightarrow0$,
together provide the regularization parameters. The techniques involved
in the Legendre polynomial expansions and the integration over $\Phi$
are described in detail in Appendices C and D of Paper I \cite{detweiler-m-w(03)}.

Below in Subsections~\ref{sub:A-terms} and \ref{sub:B-terms}, we
present the key steps of calculating the $A_{a}$ and $B_{a}$ regularization
parameters in Eqs.~(\ref{eq:10})-(\ref{eq:17}).

\subsection{$A_{a}$-terms\label{sub:A-terms}}

We take the $\epsilon^{-2}$ term from Eq.~(\ref{eq:30}) and define\begin{equation}
Q_{a}[\epsilon^{-2}]\equiv-\frac{{q^{2}}}{2}\frac{{\left.\partial_{a}\left(\tilde{{\rho}}^{2}\right)\right|_{t=t_{\mathrm{o}}}}}{\tilde{{\rho}}_{\mathrm{o}}^{3}}\label{eq:40}\end{equation}
 Then, we proceed with our calculations of the regularization parameters
one component at a time.

\subsubsection{$A_{t}$-term\emph{:}}

First we complete the expression for \textbf{$Q_{t}[\epsilon^{-2}]$}
by recalling Eqs.~(\ref{A}) and (\ref{eq:37})\textbf{\begin{eqnarray}
Q_{t}[\epsilon^{-2}] & = & -\frac{{q^{2}}}{2}\tilde{{\rho}}_{\mathrm{o}}^{-3}\left.\partial_{t}\left(\tilde{{\rho}}^{2}\right)\right|_{t=t_{\mathrm{o}}}\nonumber \\
 & = & \frac{{q^{2}}}{2}\left[2\left(r_{\mathrm{o}}^{2}+J^{2}\right)\chi\left(\delta^{2}+1-\cos\Theta\right)\right]^{-3/2}\left(\frac{{2E\dot{{r}}r_{\mathrm{o}}^{2}\Delta}}{f\left(r_{\mathrm{o}}^{2}+J^{2}\right)}+2EJ\sin\Theta\cos\Phi\right)\nonumber \\
 & = & \frac{{q^{2}E\dot{{r}}r_{\mathrm{o}}^{2}\Delta\chi^{-3/2}}}{2\sqrt{2}f\left(r_{\mathrm{o}}^{2}+J^{2}\right)^{5/2}}\left(\delta^{2}+1-\cos\Theta\right)^{-3/2}\nonumber \\
 &  & -\frac{{q^{2}EJ\chi^{-3/2}\cos\Phi}}{\sqrt{2}\left(r_{\mathrm{o}}^{2}+J^{2}\right)^{3/2}}\left.\frac{{\partial}}{\partial\Theta}\right|_{\Delta}\left(\delta^{2}+1-\cos\Theta\right)^{-1/2},\label{eq:41}\end{eqnarray}
} where $\left.\partial/\partial\Theta\right|_{\Delta}$ means that
$\Delta$ is held constant while the differentiation is performed
with respect to $\Theta$.

According to Appendix D of Paper I \cite{detweiler-m-w(03)}, for
$p\geq1$\begin{equation}
\left(\delta^{2}+1-\cos\Theta\right)^{-p-1/2}=\sum_{\ell=0}^{\infty}\frac{{2\ell+1}}{\delta^{2p-1}(2p-1)}\left[1+O(\ell\delta)\right]P_{\ell}(\cos\Theta),\;\delta\rightarrow0,\label{eq:42}\end{equation}
 and for $p=0$\begin{equation}
\left(\delta^{2}+1-\cos\Theta\right)^{-1/2}=\sum_{\ell=0}^{\infty}\left[\sqrt{2}+O(\ell\delta)\right]P_{\ell}(\cos\Theta),\;\delta\rightarrow0.\label{rho-1/2}\end{equation}
 Then, by Eqs.~(\ref{eq:42}) for $p=1$, (\ref{rho-1/2}) and (\ref{eq:39}),
in the limit $\delta\rightarrow0$ (equivalently $\Delta\rightarrow0$)
Eq.~(\ref{eq:41}) becomes

\begin{eqnarray}
\lim_{\Delta\rightarrow0}Q_{t}[\epsilon^{-2}] & = & \mathrm{sgn}(\Delta)\frac{{q^{2}\dot{{r}}r_{\mathrm{o}}\chi^{-1}}}{\left(r_{\mathrm{o}}^{2}+J^{2}\right)^{3/2}}\sum_{\ell=0}^{\infty}\left(\ell+\frac{{1}}{2}\right)P_{\ell}(\cos\Theta)\nonumber \\
 &  & -\frac{{q^{2}EJ\chi^{-3/2}\cos\Phi}}{\left(r_{\mathrm{o}}^{2}+J^{2}\right)^{3/2}}\sum_{\ell=0}^{\infty}\left.\frac{{\partial}}{\partial\Theta}\right|_{\Delta}P_{\ell}(\cos\Theta).\label{eq:43}\end{eqnarray}

Then, we integrate $\lim_{\Delta\rightarrow0}Q_{t}[\epsilon^{-2}]$
over $\Phi$ and divide it by $2\pi$ (we denote this process by the
angle brackets {}``$\langle\,\rangle$'')\begin{equation}
\left\langle \lim_{\Delta\rightarrow0}Q_{t}[\epsilon^{-2}]\right\rangle =\mathrm{sgn}(\Delta)\frac{{q^{2}\dot{{r}}r_{\mathrm{o}}\left\langle \chi^{-1}\right\rangle }}{\left(r_{\mathrm{o}}^{2}+J^{2}\right)^{3/2}}\sum_{\ell=0}^{\infty}\left(\ell+\frac{{1}}{2}\right)P_{\ell}(\cos\Theta),\label{eq:44}\end{equation}
 where we exploit the fact that $\left\langle \chi^{-3/2}\cos\Phi\right\rangle =0$
to get rid of the second term in Eq.~(\ref{eq:43}) %
\footnote{Or alternatively, one can use the argument $\left.\frac{{\partial}}{\partial\Theta}\right|_{\Delta}P_{\ell}(\cos\Theta)=0$
as $\Theta\rightarrow0$, to show that this part does not survive
at the end.%
}. Appendix C of Paper I \cite{detweiler-m-w(03)} provides $\left\langle \chi^{-1}\right\rangle ={}_{2}F_{1}\left(1,\frac{{1}}{2};1;\alpha\right)\equiv F_{1}=\left(1-\alpha\right)^{-1/2}$,
where $\alpha\equiv J^{2}/\left(r_{\mathrm{o}}^{2}+J^{2}\right)$.
Substituting this into Eq.~(\ref{eq:44}), the regularization parameter
$A_{t}$ is the coefficient of the sum on the right hand side in the
coincidence limit $\Theta\rightarrow0$\begin{equation}
A_{t}=\mathrm{sgn}(\Delta)\frac{{q^{2}}}{r_{\mathrm{o}}^{2}}\frac{{\dot{{r}}}}{1+J^{2}/r_{\mathrm{o}}^{2}}.\label{eq:45}\end{equation}

\subsubsection{$A_{r}$-term\emph{:}}

Similarly, we have\begin{equation}
Q_{r}[\epsilon^{-2}]=-\frac{{q^{2}}}{2}\tilde{{\rho}}_{\mathrm{o}}^{-3}\left.\partial_{r}\left(\tilde{{\rho}}^{2}\right)\right|_{t=t_{\mathrm{o}}}.\label{eq:46}\end{equation}
 Here, before computing $\left.\partial_{r}\left(\tilde{{\rho}}^{2}\right)\right|_{t=t_{\mathrm{o}}}$
we reverse the steps of Eqs.~(\ref{eq:33}), (\ref{qdrt}), (\ref{qdrt3})
and (\ref{A}) to obtain the relation \begin{equation}
\tilde{\rho}^{2}=\mathcal{P}_{\mathrm{II}}+O[(x-x_{\mathrm{o}})^{4}],\label{A-qdrt}\end{equation}
where $\mathcal{P}_{\mathrm{II}}$ is now back to Eq.~(\ref{eq:33}).
Differentiating this with respect to $r$ and going through the steps
of Eqs.~(\ref{qdrt}) and (\ref{eq:35}), Eq.~(\ref{eq:46}) can
be expressed with the help of Eq.~(\ref{eq:37}) as \begin{equation}
Q_{r}[\epsilon^{-2}]=-\frac{{q^{2}}}{f^{2}}\left[2\left(r_{\mathrm{o}}^{2}+J^{2}\right)\chi\left(\delta^{2}+1-\cos\Theta\right)\right]^{-3/2}\left[\frac{{r_{\mathrm{o}}^{2}E^{2}\Delta}}{r_{\mathrm{o}}^{2}+J^{2}}+fJ\dot{r}\sin\Theta\cos\Phi\right].\label{eq:Qr}\end{equation}
 Then, the rest of the calculation is carried out in the same fashion
as for the case of $A_{t}$-term above. We obtain\begin{equation}
A_{r}=-\mathrm{sgn}(\Delta)\frac{{q^{2}}}{r_{\mathrm{o}}^{2}}\frac{{E}}{f\left(1+J^{2}/r_{\mathrm{o}}^{2}\right)}.\label{eq:47}\end{equation}

\subsubsection{$A_{\phi}$-term\emph{:}}

First we have \begin{eqnarray}
Q_{\phi}[\epsilon^{-2}] & = & -\frac{{q^{2}}}{2}\tilde{{\rho}}_{\mathrm{o}}^{-3}\left.\partial_{\phi}\left(\tilde{{\rho}}^{2}\right)\right|_{t=t_{\mathrm{o}}}.\label{eq:48}\end{eqnarray}
 Taking the same steps as used for $A_{r}$-term above via Eqs.~(\ref{A-qdrt}),
(\ref{qdrt}) and (\ref{eq:35}) in order, we obtain\begin{equation}
\left.\partial_{\phi}\left(\tilde{{\rho}}^{2}\right)\right|_{t=t_{\mathrm{o}}}=2\left(r_{\mathrm{o}}^{2}+J^{2}\right)\sin\Theta\cos\Phi+O[(x-x_{\mathrm{o}})^{3}].\label{dAdphi}\end{equation}
 Then, in a similar manner to that employed in the previous cases,
in the limit $\Delta\rightarrow0$ Eq.~(\ref{eq:48}) becomes \begin{equation}
\lim_{\Delta\rightarrow0}Q_{\phi}[\epsilon^{-2}]=-\frac{{q^{2}\chi^{-3/2}\cos\Phi}}{\left(r_{\mathrm{o}}^{2}+J^{2}\right)^{1/2}}\sum_{\ell=0}^{\infty}\left.\frac{{\partial}}{\partial\Theta}\right|_{\Delta}P_{\ell}(\cos\Theta).\label{Qphi}\end{equation}
 The right hand side vanishes through {}``$\left\langle \,\right\rangle $''
process because $\left\langle \chi^{-3/2}\cos\Phi\right\rangle =0$.
Hence,\begin{equation}
A_{\phi}=0.\label{eq:49}\end{equation}

\subsubsection{$A_{\theta}$-term\emph{:} }

It is evident from the particle's motion, which is confined to the
equatorial plane $\theta_{\mathrm{o}}=\frac{{\pi}}{2}$, that no self-force
is acting on the particle in the direction perpendicular to this plane.
This is due to the fact that both the derivatives of retarded field
and the singular source field with respect to $\theta$ tend to zero
in the coincidence limit. Our calculation of $A_{\theta}$ should
support this. Through the same process as employed before, we have
\begin{equation}
Q_{\theta}[\epsilon^{-2}]=-\frac{{q^{2}}}{2}\tilde{{\rho}}_{\mathrm{o}}^{-3}\left.\partial_{\theta}\left(\tilde{{\rho}}^{2}\right)\right|_{t=t_{\mathrm{o}}}\label{eq:50}\end{equation}
 with \begin{equation}
\left.\partial_{\theta}\left(\tilde{{\rho}}^{2}\right)\right|_{t=t_{\mathrm{o}}}=2r_{\mathrm{o}}^{2}\sin\Theta\sin\Phi+O[(x-x_{\mathrm{o}})^{3}].\label{dAdtheta}\end{equation}
 Then, similarly as in the case of $A_{\phi}$-term above \begin{equation}
\lim_{\Delta\rightarrow0}Q_{\theta}[\epsilon^{-2}]=-\frac{{q^{2}r_{\mathrm{o}}^{2}\chi^{-3/2}\sin\Phi}}{\left(r_{\mathrm{o}}^{2}+J^{2}\right)^{3/2}}\sum_{\ell=0}^{\infty}\left.\frac{{\partial}}{\partial\Theta}\right|_{\Delta}P_{\ell}(\cos\Theta).\label{Qtheta}\end{equation}
 Again, via {}``$\left\langle \,\right\rangle $'' process, the
right hand side vanishes because $\left\langle \chi^{-3/2}\sin\Phi\right\rangle =0$.
Thus,\begin{equation}
A_{\theta}=0.\label{eq:51}\end{equation}

\subsection{$B_{a}$-terms\label{sub:B-terms}}

We take the $\epsilon^{-1}$ term from Eq.~(\ref{eq:30}) and define
\begin{equation}
Q_{a}[\epsilon^{-1}]\equiv q^{2}\left\{ -\frac{{1}}{2}\frac{{\left.\partial_{a}\mathcal{P}_{\mathrm{III}}\right|_{t=t_{\mathrm{o}}}}}{\tilde{{\rho}}_{\mathrm{o}}^{3}}+\frac{{3}}{4}\frac{{\left.\left[\partial_{a}\left(\tilde{{\rho}}^{2}\right)\right]\mathcal{P}_{\mathrm{III}}\right|_{t=t_{\mathrm{o}}}}}{\tilde{{\rho}}_{\mathrm{o}}^{5}}\right\} ,\label{eq:52}\end{equation}
 where for computing $\partial_{a}\left(\tilde{{\rho}}^{2}\right)$,
Eq.~(\ref{A-qdrt}) should be referred to, and $\mathcal{P}_{\mathrm{III}}$
is the cubic part taken directly from Eq.~(\ref{eq:27}).

We may express this in a generic form \begin{equation}
Q_{a}[\epsilon^{-1}]=\sum_{n=1}^{2}\sum_{k=0}^{2n}\sum_{p=0}^{[k/2]}\frac{{b_{nkp(a)}\Delta^{2n-k}\left(\phi-\phi_{\mathrm{o}}\right)^{k-2p}\left(\theta-\frac{\pi}{2}\right)^{2p}}}{\tilde{\rho}_{\mathrm{o}}^{2n+1}},\label{b1}\end{equation}
 where $\Delta\equiv r-r_{\mathrm{o}}$, and $b_{nkp(a)}$ is the
coefficient of each individual term that depends on $n$, $k$ and
$p$ as well as $a$, with a dimension $\mathcal{R}^{k-1}$ for $a=t,\, r$
and $\mathcal{R}^{k}$ for $a=\theta,\,\phi$. We recall from Eqs.~(\ref{eq:33})
and (\ref{eq:34}) that the first of the steps to lead to $\tilde{\rho}_{\mathrm{o}}^{2}$
in the denominator is replacing $\phi-\phi_{\mathrm{o}}$ by $\left(\phi-\phi'\right)-J\dot{r}\Delta/f(r_{\mathrm{o}}^{2}+J^{2})$
to eliminate the coupling term $\Delta\left(\phi-\phi_{\mathrm{o}}\right)$.
This makes a sum of independent square forms of each of $\Delta$
and $\phi-\phi'$, which is a necessary step to induce the Legendre
polynomial expansions later. Thus, to be consistent with this modification
in the denominator, $\phi-\phi_{\mathrm{o}}$ in the numerator on
the right hand side of Eq.~(\ref{b1}) should be also replaced by
$\left(\phi-\phi'\right)-J\dot{r}\Delta/f(r_{\mathrm{o}}^{2}+J^{2})$.
Then, this will create a number of additional terms apart from $\left(\phi-\phi'\right)^{m}$
when we expand the quantity $[(\phi-\phi')-J\dot{r}\Delta/f(r_{\mathrm{o}}^{2}+J^{2})]$
raised, say, to the $m$-th power, and the computation will be very
complicated.

By analyzing the structure of the quantity on the right hand side
of Eq.~(\ref{b1}) one can prove that $\phi-\phi_{\mathrm{o}}$ may
be replaced just by $\phi-\phi'$ in the numerator without the term
$-J\dot{r}\Delta/f(r_{\mathrm{o}}^{2}+J^{2})$ (the same idea is found
in Mino, Nakano, and Sasaki \cite{mino-nakano-sasaki(02)}). The verification
follows. The behavior of the quantity on the right hand side of Eq.~(\ref{b1}),
according to the powers of each factor, is \begin{equation}
Q_{a}[\epsilon^{-1}]\sim\tilde{\rho}_{\mathrm{o}}^{-(2n+1)}\Delta^{2n-k}\left(\phi-\phi_{\mathrm{o}}\right)^{k-2p}\left(\theta-\frac{\pi}{2}\right)^{2p}\mathrm{\mathcal{R}}^{s},\label{b2}\end{equation}
 where $s=k-1$ for $a=t,\, r$ and $s=k$ for $a=\theta,\,\phi$.
Further, \begin{eqnarray}
\left(\phi-\phi_{\mathrm{o}}\right)^{k-2p} & = & \left[\left(\phi-\phi'\right)-\frac{{J\dot{{r}}\Delta}}{f(r_{\mathrm{o}}^{2}+J^{2})}\right]^{k-2p}\nonumber \\
 & = & \sum_{i=0}^{k-2p}c_{kpi}\left(\phi-\phi'\right)^{i}\Delta^{k-2p-i}\sim\left(\phi-\phi'\right)^{i}\Delta^{k-2p-i}/\mathcal{R}^{k-2p-i}\label{b3}\\
 & \sim & \left(\sin\Theta\right)^{i}\left(\cos\Phi\right)^{i}\Delta^{k-2p-i}/\mathcal{R}^{k-2p-i}+O[(x-x_{\mathrm{o}})^{k-2p+2}],\label{b4}\end{eqnarray}
 where a binomial expansion over the index $i=0,\,1,\,\cdots\,,\, k-2p$
is assumed with $c_{kpi}\sim1/\mathcal{R}^{k-2p-i}$ in Eq.~(\ref{b3}),
and in Eq.~(\ref{b4}) $\left(\phi-\phi'\right)^{i}$ is replaced
by $[\sin(\phi-\phi')]^{i}+O[\left(\phi-\phi'\right)^{i+2}]$ ---
the term $O[(x-x_{\mathrm{o}})^{k-2p+2}]$ at the end results from
this $O[\left(\phi-\phi'\right)^{i+2}]$, then the coordinates are
rotated using the definition of new angles by Eq.~(\ref{eq:35}).
Also, by Eq.~(\ref{eq:35}) again \begin{equation}
\left(\theta-\frac{\pi}{2}\right)^{2p}=\left(\sin\Theta\right)^{2p}\left(\sin\Phi\right)^{2p}+O[(x-x_{\mathrm{o}})^{2p+2}].\label{b5}\end{equation}
 Using Eqs.~(\ref{b4}) and (\ref{b5}), the behavior of $Q[\epsilon^{-1}]$
in Eq.~(\ref{b2}) looks like \begin{equation}
Q_{a}[\epsilon^{-1}]\sim\tilde{\rho}_{\mathrm{o}}^{-(2n+1)}\Delta^{2n-2p-i}\left(\sin\Theta\right)^{2p+i}\left(\cos\Phi\right)^{i}\left(\sin\Phi\right)^{2p}\mathcal{R}^{s},\label{b6}\end{equation}
 where $s=2p+i-1$ for $a=t,\, r$ and $s=2p+i$ for $a=\theta,\,\phi$,
and any contributions from $O[(x-x_{\mathrm{o}})^{k-2p+2}]$ in Eq.~(\ref{b4})
and from $O[(x-x_{\mathrm{o}})^{2p+2}]$ in Eq.~(\ref{b5}) have
been disregarded: by putting these pieces into Eq.~(\ref{b2}) we
simply obtain $\epsilon^{1}$-terms, which would correspond to $O(\ell^{-2})$
in Eq.~(\ref{eq:9}) and should vanish when summed over $\ell$ in
our final self-force calculation by Eq.~(\ref{eq:8}). $Q_{a}[\epsilon^{-1}]$
then can be categorized into the following cases:

\begin{enumerate}
\item $i=2j+1$ ($j=0,\,1,\,2,\,\cdots$) \\
 The integrand for {}``$\left\langle \,\right\rangle $'' process,
$F(\Phi)\equiv\left(\cos\Phi\right)^{2j+1}\left(\sin\Phi\right)^{2p}$
has the property $F(\Phi+\pi)=-F(\Phi)$. Thus \begin{equation}
\left\langle Q_{a}[\epsilon^{-1}]\right\rangle =0,\label{b7}\end{equation}

\item $i=2j$ ($j=0,\,1,\,2,\,\cdots$)\\
 Using Eqs.~(\ref{eq:37}) and (\ref{eq:39}), we can express $\left(\sin\Theta\right)^{2p+i}$
in Eq.~(\ref{b6}) above in terms of $\tilde{\rho}_{\mathrm{o}}$
and $\Delta$ via a binomial expansion \begin{eqnarray}
\left(\sin\Theta\right)^{2p+2j} & = & \left[2\left(1-\cos\Theta\right)\right]^{p+j}+O[(x-x_{\mathrm{o}})^{2(p+j)+2}]\nonumber \\
 & = & \sum_{q=0}^{p+j}d_{pjq}\tilde{\rho}_{\mathrm{o}}^{2q}\Delta^{2(p+j-q)}+O[(x-x_{\mathrm{o}})^{2(p+j)+2}]\\
 & \sim & \tilde{\rho}_{\mathrm{o}}^{2q}\Delta^{2(p+j-q)}/\mathcal{R}^{2(p+j)}+O[(x-x_{\mathrm{o}})^{2(p+j)+2}],\label{b13}\end{eqnarray}
 where $q=0,\,1,\,\cdots\,,\, p+j$ is the index for a binomial expansion
and $d_{pjq}\sim1/\mathcal{R}^{2(p+j)}$. When Eq.~(\ref{b13}) is
substituted into Eq.~(\ref{b6}), the contribution from $O[(x-x_{\mathrm{o}})^{2(p+j)+2}]$
can be disregarded since it would correspond to $O(\epsilon^{1})$
again. Then, we have \begin{equation}
Q_{a}[\epsilon^{-1}]\sim\left(\sin\Phi\right)^{2p}\left(\cos\Phi\right)^{2j}\tilde{\rho}_{\mathrm{o}}^{-2(n-q)-1}\Delta^{2(n-q)}\mathcal{R}^{s},\label{b8}\end{equation}
 where $s=-1$ for $a=t,\, r$ and $s=0$ for $a=\theta,\,\phi$,
and we can guarantee that $n-q\geq0$ always since $0\leq q\leq p+j=p+\frac{{1}}{2}i$,
$0\leq i\leq k-2p$ and $p\leq k\leq2n$. Then, Eq.~(\ref{b8}) can
be subcategorized into the following two cases;

\begin{enumerate}
\item $n-q\geq1$\\
 By Eqs.~(\ref{eq:37}), (\ref{eq:39}) and (\ref{eq:42}) \begin{equation}
Q_{a}[\epsilon^{-1}]\begin{array}{c}
\\\sim\\
^{\Delta\rightarrow0}\end{array}\left(\sin\Phi\right)^{2p}\left(\cos\Phi\right)^{2j}\Delta P_{\ell}(\cos\Theta)\mathcal{R}^{s}\longrightarrow0,\label{b9}\end{equation}

\item $n-q=0$\\
 By Eqs.~(\ref{eq:37}), (\ref{eq:39}) and (\ref{rho-1/2}) \begin{equation}
Q_{a}[\epsilon^{-1}]\begin{array}{c}
\\\sim\\
^{\Delta\rightarrow0}\end{array}\left(\sin\Phi\right)^{2p}\left(\cos\Phi\right)^{2j}P_{\ell}(\cos\Theta)\mathcal{R}^{s},\label{b10}\end{equation}
 where $s=-1$ for $a=t,\, r$ and $s=0$ for $a=\theta,\,\phi$.
\end{enumerate}
\end{enumerate}
Therefore, by analyzing the structure of $Q_{a}[\epsilon^{-1}]$ we
find that the $\epsilon^{-1}$-terms vanish in all the cases except
when $n-q=0$. The non-vanishing $B_{a}$-terms are derived only from
this case. Then, by $0\leq q\leq p+j=p+\frac{{1}}{2}i$, $0\leq i\leq k-2p$
and $p\leq k\leq2n$ together with $n=q$ one can show that\begin{eqnarray}
0 & \leq & k-2p-i\;\mathrm{and}\; k-2p-i\leq0,\;\mathrm{i.e}.\; k-2p-i=0.\label{b11}\end{eqnarray}
 Substituting this result into Eq.~(\ref{b3}), then into Eq.~(\ref{b1})
we may conclude that in the numerator of $Q[\epsilon^{-1}]$ in Eq.~(\ref{b1})
one can simply substitute \begin{equation}
\left(\phi-\phi_{\mathrm{o}}\right)^{k-2p}\rightarrow\left(\phi-\phi'\right)^{k-2p}.\;\mathrm{Q}.\,\mathrm{E.}\,\mathrm{D}.\label{b12}\end{equation}

The significance of this proof does not lie in the result given by
Eq.~(\ref{b12}) only, but also in the fact that the non-vanishing
contribution comes only from the case $n=q$ for Eq.~(\ref{b8}),
i.e.\begin{equation}
Q_{a}[\epsilon^{-1}]\sim\left(\sin\Phi\right)^{2p}\left(\cos\Phi\right)^{2(n-p)}\tilde{\rho}_{\mathrm{o}}^{-1}\mathcal{R}^{s},\label{b14}\end{equation}
 where $n=1,\,2$ and $0\leq p\leq n$, and $s=-1$ for $a=t,\, r$
and $s=0$ for $a=\theta,\,\phi$.

Below are presented the calculations of $B_{a}$-terms of the regularization
parameters by component, in a similar manner to those for $A_{a}$-terms.

\subsubsection{$B_{t}$-term\emph{:}}

We begin with\begin{equation}
Q_{t}[\epsilon^{-1}]=q^{2}\left\{ -\frac{{1}}{2}\frac{{\left.\partial_{t}\mathcal{P}_{\mathrm{III}}\right|_{t=t_{\mathrm{o}}}}}{\tilde{{\rho}}_{\mathrm{o}}^{3}}+\frac{{3}}{4}\frac{{\left.\left[\partial_{t}\left(\tilde{{\rho}}^{2}\right)\right]\mathcal{P}_{\mathrm{III}}\right|_{t=t_{\mathrm{o}}}}}{\tilde{{\rho}}_{\mathrm{o}}^{5}}\right\} .\label{eq:53}\end{equation}
 The subsequent computation will be very lengthy and it will be reasonable
to split $Q_{t}[\epsilon^{-1}]$ into two parts. First, let \begin{equation}
Q_{t(1)}[\epsilon^{-1}]\equiv-\frac{{q^{2}}}{2}\tilde{{\rho}}_{\mathrm{o}}^{-3}\left.\partial_{t}\mathcal{P}_{\mathrm{III}}\right|_{t=t_{\mathrm{o}}},\label{eq:56}\end{equation}
 where\begin{equation}
\left.\partial_{t}\mathcal{P}_{\mathrm{III}}\right|_{t=t_{\mathrm{o}}}=-\frac{{ME\dot{{r}}\Delta^{2}}}{f^{2}r_{\mathrm{o}}^{2}}-2\left(1-\frac{{M}}{r_{\mathrm{o}}}\right)\frac{{EJ\Delta}}{fr_{\mathrm{o}}}\left(\phi-\phi_{\mathrm{o}}\right)+r_{\mathrm{o}}E\dot{{r}}\left[\left(\phi-\phi_{\mathrm{o}}\right)^{2}+\left(\theta-\frac{{\pi}}{2}\right)^{2}\right].\label{eq:54}\end{equation}
 As proved at the beginning of this Subsection, every $\left(\phi-\phi_{\mathrm{o}}\right)^{m}$
in the numerators of the $\epsilon^{-1}$-term can be replaced by
$\left(\phi-\phi'\right)^{m}$ without affecting the rest of calculation.
Then, followed by the rotation of the coordinates via Eq.~(\ref{eq:35})
\begin{eqnarray}
Q_{t(1)}[\epsilon^{-1}] & = & -\frac{{q^{2}}}{2}\tilde{{\rho}}_{\mathrm{o}}^{-3}\left[-\frac{{ME\dot{{r}}\Delta^{2}}}{f^{2}r_{\mathrm{o}}^{2}}-2\left(1-\frac{{M}}{r_{\mathrm{o}}}\right)\frac{{EJ\Delta}}{fr_{\mathrm{o}}}\sin\Theta\cos\Phi+2r_{\mathrm{o}}E\dot{{r}}\left(1-\cos\Theta\right)\right]\nonumber \\
 &  & +O\left[\frac{{(x-x_{\mathrm{o}})^{4}}}{\tilde{{\rho}}_{\mathrm{o}}^{3}}\right],\label{eq:57}\end{eqnarray}
 where an approximation $\sin^{2}\Theta=2(1-\cos\Theta)+O[(x-x_{\mathrm{o}})^{4}]$
is used to obtain the last term inside the first bracket. Here we
may drop off the term $O\left[(x-x_{\mathrm{o}})^{4}/\tilde{{\rho}}_{\mathrm{o}}^{3}\right]$
, which is essentially $O(\epsilon^{1})$, for the same reason as
explained at the beginning of this subsection. Then, using the same
techniques as used to find $A_{a}$-terms, we can reduce Eq.~(\ref{eq:57})
to\begin{eqnarray}
Q_{t(1)}[\epsilon^{-1}] & = & \left[\frac{{q^{2}ME\dot{{r}}}}{2f^{2}r_{\mathrm{o}}^{2}}+\frac{{q^{2}r_{\mathrm{o}}^{3}E^{3}\dot{{r}}\chi^{-1}}}{2f^{2}\left(r_{\mathrm{o}}^{2}+J^{2}\right)^{2}}\right]\Delta^{2}\left[2\left(r_{\mathrm{o}}^{2}+J^{2}\right)\chi\left(\delta^{2}+1-\cos\Theta\right)\right]^{-3/2}\nonumber \\
 &  & -\frac{{q^{2}\left(1-\frac{{M}}{r_{\mathrm{o}}}\right)EJ\Delta\chi^{-3/2}\cos\Phi}}{\sqrt{2}fr_{\mathrm{o}}\left(r_{\mathrm{o}}^{2}+J^{2}\right)^{3/2}}\left.\frac{{\partial}}{\partial\Theta}\right|_{\Delta}\left(\delta^{2}+1-\cos\Theta\right)^{-1/2}\nonumber \\
 &  & -\frac{{q^{2}E\dot{{r}}r_{\mathrm{o}}\chi^{-1}}}{2\left(r_{\mathrm{o}}^{2}+J^{2}\right)}\tilde{{\rho}}_{\mathrm{o}}^{-1}.\label{eq:58}\end{eqnarray}
 As we have seen before, by Eq.~(\ref{eq:42}) $\left(\delta^{2}+1-\cos\Theta\right)^{-3/2}\sim\Delta^{-1}$
in the limit $\Delta\rightarrow0$ and the first term on the right
hand side will vanish. The second term will also give no contribution
to the regularization parameters because $\left\langle \chi^{-3/2}\cos\Phi\right\rangle =0$.
Only the last term, which is $\sim\tilde{{\rho}}_{\mathrm{o}}^{-1}$,
will give non-zero contribution according to the argument in the analysis
presented above (see Eq.~(\ref{b14})). Using Eq.~(\ref{rho-1/2})
in the limit $\Delta\rightarrow0$ and taking {}``$\left\langle \,\right\rangle $''
process, Eq.~(\ref{eq:58}) becomes\begin{equation}
\left\langle \lim_{\Delta\rightarrow0}Q_{t(1)}[\epsilon^{-1}]\right\rangle =-\frac{{1}}{2}\frac{{q^{2}}}{r_{\mathrm{o}}^{2}}\frac{{E\dot{{r}}\left\langle \chi^{-3/2}\right\rangle }}{\left(1+J^{2}/r_{\mathrm{o}}^{2}\right)^{3/2}}\sum_{\ell=0}^{\infty}P_{\ell}\left(\cos\Theta\right).\label{eq:59}\end{equation}
 The identity $\left\langle \chi^{-p}\right\rangle \equiv\left\langle \left(1-\alpha\sin^{2}\Phi\right)^{-p}\right\rangle ={}_{2}F_{1}\left(p,\,\frac{{1}}{2};\,1,\,\alpha\right)\equiv F_{p}$,
with $\alpha\equiv J^{2}/\left(r_{\mathrm{o}}^{2}+J^{2}\right)$ is
taken from Appendix C of Paper I \cite{detweiler-m-w(03)}, and we
take the limit $\Theta\rightarrow0$\begin{equation}
\left.\left\langle \lim_{\Delta\rightarrow0}Q_{t(1)}[\epsilon^{-1}]\right\rangle \right|_{\Theta\rightarrow0}=-\frac{{1}}{2}\frac{{q^{2}}}{r_{\mathrm{o}}^{2}}\frac{{E\dot{{r}}F_{3/2}}}{\left(1+J^{2}/r_{\mathrm{o}}^{2}\right)^{3/2}}.\label{eq:61}\end{equation}

Now the remaining part is\begin{equation}
Q_{t(2)}[\epsilon^{-1}]\equiv\frac{{3q^{2}}}{4}\tilde{{\rho}}_{\mathrm{o}}^{-5}\left.\left[\partial_{t}\left(\tilde{{\rho}}^{2}\right)\right]\mathcal{P}_{\mathrm{III}}\right|_{t=t_{\mathrm{o}}},\label{eq:62}\end{equation}
 where

\begin{eqnarray}
\left.\left[\partial_{t}\left(\tilde{{\rho}}^{2}\right)\right]\mathcal{P}_{\mathrm{III}}\right|_{t=t_{\mathrm{o}}} & = & \left[-\frac{{2E\dot{{r}}\Delta}}{f}-2EJ\left(\phi-\phi_{\mathrm{o}}\right)\right]\nonumber \\
 &  & \times\left[-\left(1+\frac{{\dot{{r}}^{2}}}{f}\right)\frac{{M\Delta^{3}}}{f^{2}r_{\mathrm{o}}^{2}}+\left(2-\frac{{5M}}{r_{\mathrm{o}}}\right)\frac{{J\dot{{r}}\Delta^{2}}}{f^{2}r_{\mathrm{o}}}\left(\phi-\phi_{\mathrm{o}}\right)\right.\nonumber \\
 &  & +\left(1-\frac{{\dot{{r}}}}{f}+\frac{{2J^{2}}}{r_{\mathrm{o}}^{2}}\right)r_{\mathrm{o}}\Delta\left(\phi-\phi_{\mathrm{o}}\right)^{2}+\left(1-\frac{{\dot{{r}}^{2}}}{f}\right)r_{\mathrm{o}}\Delta\left(\theta-\frac{{\pi}}{2}\right)^{2}\nonumber \\
 &  & \left.-r_{\mathrm{o}}J\dot{{r}}\left(\phi-\phi_{\mathrm{o}}\right)^{3}-r_{\mathrm{o}}J\dot{{r}}\left(\phi-\phi_{\mathrm{o}}\right)\left(\theta-\frac{{\pi}}{2}\right)^{2}\right]+O[(x-x_{\mathrm{o}})^{6}].\label{eq:55}\end{eqnarray}
 Taking similar procedures as above, the non-vanishing contributions
turn out to be\begin{eqnarray}
\left\langle \lim_{\Delta\rightarrow0}Q_{t(2)}[\epsilon^{-1}]\right\rangle  & = & \left\langle \lim_{\Delta\rightarrow0}\frac{{3}}{2}q^{2}EJ^{2}\dot{{r}}r_{\mathrm{o}}\tilde{{\rho}}_{\mathrm{o}}^{-5}\cos^{2}\Phi\sin^{4}\Theta\right\rangle \nonumber \\
 & = & \left\langle \lim_{\Delta\rightarrow0}\frac{{3}}{2}\frac{{q^{2}}}{r_{\mathrm{o}}}\frac{{E\dot{{r}}\tilde{{\rho}}_{\mathrm{o}}^{-1}}}{1+J^{2}/r_{\mathrm{o}}^{2}}\left(\chi^{-1}-\frac{{\chi^{-2}}}{1+J^{2}/r_{\mathrm{o}}^{2}}\right)\right\rangle \nonumber \\
 & = & \frac{{3}}{2}\frac{{q^{2}}}{r_{\mathrm{o}}^{2}}\frac{{E\dot{{r}}}}{\left(1+J^{2}/r_{\mathrm{o}}^{2}\right)^{3/2}}\left(\left\langle \chi^{-3/2}\right\rangle -\frac{{\left\langle \chi^{-5/2}\right\rangle }}{1+J^{2}/r_{\mathrm{o}}^{2}}\right)\sum_{\ell=0}^{\infty}P_{\ell}\left(\cos\Theta\right),\label{eq:63}\end{eqnarray}
 where all other terms than $\sim\tilde{{\rho}}_{\mathrm{o}}^{-1}$
again have been dropped off during the procedure since they vanish
either in the limit $\Delta\rightarrow0$ or through the {}``$\left\langle \,\right\rangle $''
process. Then, using the identity $\left\langle \chi^{-p}\right\rangle \equiv{}_{2}F_{1}\left(p,\,\frac{{1}}{2};\,1,\,\alpha\right)\equiv F_{p}$,
we have\begin{equation}
\left.\left\langle \lim_{\Delta\rightarrow0}Q_{t(2)}[\epsilon^{-1}]\right\rangle \right|_{\Theta\rightarrow0}=\frac{{3}}{2}\frac{{q^{2}}}{r_{\mathrm{o}}^{2}}\frac{{E\dot{{r}}}}{\left(1+J^{2}/r_{\mathrm{o}}^{2}\right)^{3/2}}\left(F_{3/2}-\frac{{F_{5/2}}}{1+J^{2}/r_{\mathrm{o}}^{2}}\right).\label{eq:65}\end{equation}

By combining Eqs.~(\ref{eq:61}) and (\ref{eq:65}), we finally obtain\begin{equation}
B_{t}=\frac{{q^{2}}}{r_{\mathrm{o}}^{2}}E\dot{r}\left[\frac{{F_{3/2}}}{\left(1+J^{2}/r_{\mathrm{o}}^{2}\right)^{3/2}}-\frac{{3F_{5/2}}}{2\left(1+J^{2}/r_{\mathrm{o}}^{2}\right)^{5/2}}\right].\label{eq:66}\end{equation}

\subsubsection{$B_{r}$-term\emph{:}}

From Eq.~(\ref{eq:52}) we start with\begin{equation}
Q_{r}[\epsilon^{-1}]=q^{2}\left\{ -\frac{{1}}{2}\frac{{\left.\partial_{r}\mathcal{P}_{\mathrm{III}}\right|_{t=t_{\mathrm{o}}}}}{\tilde{{\rho}}_{\mathrm{o}}^{3}}+\frac{{3}}{4}\frac{{\left.\left[\partial_{r}\left(\tilde{{\rho}}^{2}\right)\right]\mathcal{P}_{\mathrm{III}}\right|_{t=t_{\mathrm{o}}}}}{\tilde{{\rho}}_{\mathrm{o}}^{5}}\right\} .\label{eq:67}\end{equation}
 Then, following the same steps as taken for the case of $B_{t}$-term
above, we obtain\begin{equation}
B_{r}=\frac{{q^{2}}}{r_{\mathrm{o}}^{2}}\left[-\frac{{F_{1/2}}}{\left(1+J^{2}/r_{\mathrm{o}}^{2}\right)^{1/2}}+\frac{{(1-2f^{-1}\dot{{r}}^{2}){F_{3/2}}}}{2\left(1+J^{2}/r_{\mathrm{o}}^{2}\right)^{3/2}}+\frac{{3}f^{-1}\dot{{r}}^{2}F_{5/2}}{2\left(1+J^{2}/r_{\mathrm{o}}^{2}\right)^{5/2}}\right].\label{eq:68}\end{equation}

\subsubsection{$B_{\phi}$-term\emph{:}}

Again, from Eq.~(\ref{eq:52})\begin{equation}
Q_{\phi}[\epsilon^{-1}]=q^{2}\left\{ -\frac{{1}}{2}\frac{{\left.\partial_{\phi}\mathcal{P}_{\mathrm{III}}\right|_{t=t_{\mathrm{o}}}}}{\tilde{{\rho}}_{\mathrm{o}}^{3}}+\frac{{3}}{4}\frac{{\left.\left[\partial_{\phi}\left(\tilde{{\rho}}^{2}\right)\right]\mathcal{P}_{\mathrm{III}}\right|_{t=t_{\mathrm{o}}}}}{\tilde{{\rho}}_{\mathrm{o}}^{5}}\right\} .\label{eq:69}\end{equation}
 Then, similarly we can derive\begin{equation}
B_{\phi}=\frac{{q^{2}}}{J}\dot{r}\left[\frac{{F_{1/2}-F_{3/2}}}{\left(1+J^{2}/r_{\mathrm{o}}^{2}\right)^{1/2}}+\frac{{3(F_{5/2}-F_{3/2})}}{2\left(1+J^{2}/r_{\mathrm{o}}^{2}\right)^{3/2}}\right].\label{eq:70}\end{equation}

\subsubsection{$B_{\theta}$-term\emph{:}}

As $A_{\theta}$ vanishes, so should $B_{\theta}$. From \begin{equation}
Q_{\theta}[\epsilon^{-1}]=q^{2}\left\{ -\frac{{1}}{2}\frac{{\left.\partial_{\theta}\mathcal{P}_{\mathrm{III}}\right|_{t=t_{\mathrm{o}}}}}{\tilde{{\rho}}_{\mathrm{o}}^{3}}+\frac{{3}}{4}\frac{{\left.\left[\partial_{\theta}\left(\tilde{{\rho}}^{2}\right)\right]\mathcal{P}_{\mathrm{III}}\right|_{t=t_{\mathrm{o}}}}}{\tilde{{\rho}}_{\mathrm{o}}^{5}}\right\} ,\label{eq:71}\end{equation}
 one finds that there is no term like $\sim\tilde{{\rho}}_{\mathrm{o}}^{-1}$:
all terms are either like $\sim\Delta^{2n}/\tilde{{\rho}}_{\mathrm{o}}^{2n+1}$
or like $\sim\Delta^{2n-1}\sin\Theta\cos\Phi/\tilde{{\rho}}_{\mathrm{o}}^{2n+1}$
($n=1,\,2$), which vanish in the limit $\Delta\rightarrow0$ or through
the {}``$\left\langle \,\right\rangle $'' process. Thus\begin{equation}
B_{\theta}=0.\label{eq:72}\end{equation}

\subsection{$C_{a}$-terms\label{sub:C-terms}}

We have mentioned before that $C_{a}$-terms, which originate from
$\epsilon^{0}$-term in Eq.~(\ref{eq:30}), always vanish. This can
be proved by analyzing the structure of $\epsilon^{0}$-term. First
we specify the $\epsilon^{0}$-order term for $\left.\partial_{a}(1/\rho)\right|_{t=t_{\mathrm{o}}}$
in a Laurent series expansion and define \begin{eqnarray}
Q_{a}[\epsilon^{0}] & \equiv & q^{2}\left\{ -\frac{{1}}{2}\frac{{\left.\partial_{a}\mathcal{P}_{\mathrm{IV}}\right|_{t=t_{\mathrm{o}}}}}{\tilde{{\rho}}_{\mathrm{o}}^{3}}+\frac{{3}}{4}\frac{{\left.\left(\partial_{a}\mathcal{P}_{\mathrm{III}}\right)\mathcal{P}_{\mathrm{III}}\right|_{t=t_{\mathrm{o}}}+\left.\left[\partial_{a}\left(\tilde{{\rho}}^{2}\right)\right]\mathcal{P}_{\mathrm{IV}}\right|_{t=t_{\mathrm{o}}}}}{\tilde{{\rho}}_{\mathrm{o}}^{5}}\right.\nonumber \\
 &  & \left.-\frac{{15}}{16}\frac{{\left.\left[\partial_{a}\left(\tilde{{\rho}}^{2}\right)\right]\mathcal{P}_{\mathrm{III}}^{2}\right|_{t=t_{\mathrm{o}}}}}{\tilde{{\rho}}_{\mathrm{o}}^{7}}\right\} .\label{eq:c0}\end{eqnarray}
 Generically, this can be written as

\begin{equation}
Q_{a}[\epsilon^{0}]=\sum_{n=1}^{3}\sum_{k=0}^{2n+1}\sum_{p=0}^{[k/2]}\frac{{c_{nkp(a)}\Delta^{2n+1-k}\left(\phi-\phi_{\mathrm{o}}\right)^{k-2p}\left(\theta-\frac{\pi}{2}\right)^{2p}}}{\tilde{\rho}_{\mathrm{o}}^{2n+1}},\label{c1}\end{equation}
 where $\Delta\equiv r-r_{\mathrm{o}}$, and $c_{nkp(a)}$ is the
coefficient of each individual term that depends on $n$, $k$ and
$p$ as well as $a$, with a dimension $\mathcal{R}^{k-2}$ for $a=t,\, r$
and $\mathcal{R}^{k-1}$ for $a=\theta,\,\phi$.

The behavior of $Q_{a}[\epsilon^{0}]$, according to the powers of
each factor on the right hand side of Eq.~(\ref{c1}), is \begin{equation}
Q_{a}[\epsilon^{0}]\sim\tilde{\rho}_{\mathrm{o}}^{-(2n+1)}\Delta^{2n+1-k}\left(\phi-\phi_{\mathrm{o}}\right)^{k-2p}\left(\theta-\frac{\pi}{2}\right)^{2p}\mathcal{R}^{s},\label{c2}\end{equation}
 where $s=k-2$ for $a=t,\, r$ and $s=k-1$ for $a=\theta,\,\phi$.
Following the same procedure as in the beginning of Subsection~\ref{sub:B-terms},
Eq.~(\ref{c2}) becomes

\begin{equation}
Q_{a}[\epsilon^{0}]\sim\tilde{\rho}_{\mathrm{o}}^{-(2n+1)}\Delta^{2n+1-2p-i}\left(\sin\Theta\right)^{2p+i}\left(\sin\Phi\right)^{2p}\left(\cos\Phi\right)^{i}\mathcal{R}^{s},\label{c3}\end{equation}
 where a binomial expansion over the index $i=0,\,1,\,\cdots\,,\, k-2p$
is assumed, and $s=2p+i-2$ for $a=t,\, r$ and $s=2p+i-1$ for $a=\theta,\,\phi$.
Here we have disregarded any by-products like $O[(x-x_{\mathrm{o}})^{k-2p+2}]$
and $O[(x-x_{\mathrm{o}})^{2p+2}]$, which originate from $\left(\phi-\phi_{\mathrm{o}}\right)^{k-2p}$
and $\left(\theta-\frac{\pi}{2}\right)^{2p}$, respectively when we
rotate the angles: by putting them back into Eq.~(\ref{c2}) we simply
obtain $\epsilon^{2}$-terms, which would correspond to $O(\ell^{-4})$
in Eq.~(\ref{eq:9}) and should vanish when summed over $\ell$ in
our final self-force calculation by Eq.~(\ref{eq:8}). Then, the
rest of the argument is developed in the same way as in the beginning
of Subsection~\ref{sub:B-terms}:

\begin{enumerate}
\item $i=2j+1$ ($j=0,\,1,\,2,\,\cdots$) \\
 The integrand for {}``$\left\langle \,\right\rangle $'' process,
$F(\Phi)\equiv\left(\cos\Phi\right)^{2j+1}\left(\sin\Phi\right)^{2p}$
has the property $F(\Phi+\pi)=-F(\Phi)$. Thus \begin{equation}
\left\langle Q_{a}[\epsilon^{0}]\right\rangle =0,\label{c4}\end{equation}

\item $i=2j$ ($j=0,\,1,\,2,\,\cdots$) \\
 We have \begin{equation}
Q_{a}[\epsilon^{0}]\sim\left(\sin\Phi\right)^{2p}\left(\cos\Phi\right)^{2j}\tilde{\rho}_{\mathrm{o}}^{-2(n-q)-1}\Delta^{2(n-q)+1}\mathcal{R}^{s},\label{c5}\end{equation}
 where $q=0,\,1,\,\cdots\,,\, p+j$ is the index for a binomial expansion
and $s=-2$ for $a=t,\, r$ and $s=-1$ for $a=\theta,\,\phi$. Here
we can guarantee that $n-q\geq-\frac{{1}}{2}$, i.e. $n-q=0,\,1,\,2,\,\cdots$
since $0\leq q\leq p+j=p+\frac{{1}}{2}i$, $0\leq i\leq k-2p$ and
$p\leq k\leq2n+1$. Then, Eq.~(\ref{c5}) can be subcategorized into
the following two cases;

\begin{enumerate}
\item $n-q\geq1$\\
 By Eqs.~(\ref{eq:37}), (\ref{eq:39}) and (\ref{eq:42}) \begin{equation}
Q_{a}[\epsilon^{0}]\begin{array}{c}
\\\sim\\
^{\Delta\rightarrow0}\end{array}\left(\sin\Phi\right)^{2p}\left(\cos\Phi\right)^{2j}\Delta^{2}P_{\ell}(\cos\Theta)\mathcal{R}^{s}\longrightarrow0,\label{c6}\end{equation}

\item $n-q=0$\\
 By Eqs.~(\ref{eq:37}), (\ref{eq:39}) and (\ref{rho-1/2}) \begin{equation}
Q_{a}[\epsilon^{0}]\begin{array}{c}
\\\sim\\
^{\Delta\rightarrow0}\end{array}\left(\sin\Phi\right)^{2p}\left(\cos\Phi\right)^{2j}\Delta P_{\ell}(\cos\Theta)\mathcal{R}^{s}\longrightarrow0,\label{c7}\end{equation}
 where $s=-2$ for $a=t,\, r$ and $s=-1$ for $a=\theta,\,\phi$.
\end{enumerate}
\end{enumerate}
Clearly, in any cases the quantity $Q_{a}[\epsilon^{0}]$ does not
survive, therefore we can conclude that $C_{a}$-terms are always
zero.~Q. E. D.

Also, this justifies the argument that we need not clarify the term
$O[(x-x_{\mathrm{o}})^{3}]$ in Eq.~(\ref{eq:20}) and its contribution
to $\rho^{2}$, which is $O[(x-x_{\mathrm{o}})^{4}]$ in Eqs.~(\ref{eq:26})
and (\ref{eq:27}) in Section \ref{sec:DETERMINATION-OF-psiS} or
$\mathcal{P}_{\mathrm{IV}}$ in Eqs.~(\ref{eq:29}) and (\ref{newrho2})
in Section \ref{sec:-REGULARIZATION-PARAMETERS}: by the analysis
of the generic structure given above, $-\frac{{1}}{2}\left.\partial_{a}\mathcal{P}_{\mathrm{IV}}\right|_{t=t_{\mathrm{o}}}/\tilde{{\rho}}_{\mathrm{o}}^{3}$
or $\frac{{3}}{4}\left.\left[\partial_{a}\left(\tilde{{\rho}}^{2}\right)\right]\mathcal{P}_{\mathrm{IV}}\right|_{t=t_{\mathrm{o}}}/\tilde{{\rho}}_{\mathrm{o}}^{5}$
would simply vanish in the coincidence limit $x\rightarrow x_{\mathrm{o}}$,
regardless of what $\mathcal{P}_{\mathrm{IV}}$ is.

\section{DISCUSSION}

Self-force analysis in curved spacetime relies upon the ability to
divide the field of a point charge into two parts. One part is singular
and exerts no net force on the charge itself. The remainder is a regular,
smooth vacuum field and is entirely the cause of any self-force. We
see, in this manuscript, that the singular field is adequately approximated
by its Coulomb field in a coordinate system which is locally inertial
and centered upon the charge. With this elementary approximation of
the singular field, it is guaranteed that the remainder is at least
differentiable and provides the correct self-force. For a charge moving
in the Schwarzschild geometry, the multipole moments of the singular
field are the regularization parameters which are necessary for computing
the self-force from a multipole expansion, and these regularization
parameters are calculated with a relatively elementary procedure.
Our analysis agrees with that of others \cite{barack-ori(02),mino-nakano-sasaki(02)}
and appears to us to be the most straightforward calculation of these
important parameters.

Future work will use a higher order approximation for the singular
field which will result in a more accurate and more differentiable
approximation for the regular remainder. In practice, a higher order
approximation of the self-force significantly speeds up convergence
in a mode sum, as demonstrated in Paper I \cite{detweiler-m-w(03)}.

The simplicity of our methods should also make them useful for self-force
analysis in the context of the Kerr geometry.

Another avenue for future work involves calculating gravitational
self-force regularization parameters. Thus far, published parameters
\cite{barack-ori(02),mino-nakano-sasaki(02)} focus upon the self-force,
rather than upon the the metric perturbations themselves. However,
it is clear that a gravitational self-force calculation of a gauge
invariant quantity requires the metric perturbations explicitly as
well.

Future work will find the THZ coordinates to higher orders, and calculate
higher order regularization parameters which will provide faster convergence
of the $\ell$ sums and correspondingly more accurate results.

\begin{acknowledgments}
I would like to thank Professor Steven Detweiler for many helpful
suggestions and stimulating discussions during the course of this
project and in preparing this manuscript. I would also like to thank
Professor Bernard Whiting and Professor Richard Woodard for many useful
discussions.
\end{acknowledgments}
\appendix

\section{HYPERGEOMETRIC FUNCTIONS AND REPRESENTATIONS OF REGULARIZATION PARAMETERS\label{hyper}}

In Section \ref{sec:-REGULARIZATION-PARAMETERS} we define

\begin{equation}
\chi\equiv1-\alpha\sin^{2}\Phi\label{C1}\end{equation}
 with

\begin{equation}
\alpha\equiv\frac{{J^{2}}}{r_{\mathrm{o}}^{2}+J^{2}}.\label{C2}\end{equation}
 And we use\begin{eqnarray}
\left\langle \chi^{-p}\right\rangle \equiv\left\langle \left(1-\alpha\sin^{2}\Phi\right)^{-p}\right\rangle  & = & \frac{{2}}{\pi}\int_{0}^{\pi/2}\left(1-\alpha\sin^{2}\Phi\right)^{-p}d\Phi\nonumber \\
 & = & {}_{2}F_{1}\left(p,\frac{{1}}{2};1,\alpha\right)\equiv F_{p}.\label{C3}\end{eqnarray}
 In particular, for the cases $p=\frac{{1}}{2}$ and $p=-\frac{{1}}{2}$
we have the following representations\begin{equation}
F_{1/2}={}_{2}F_{1}\left(\frac{{1}}{2},\frac{{1}}{2},1;\alpha\right)=\frac{{2}}{\pi}\hat{K}(\alpha)\label{C4}\end{equation}
 and\begin{equation}
F_{-1/2}={}_{2}F_{1}\left(-\frac{{1}}{2},\frac{{1}}{2},1;\alpha\right)=\frac{{2}}{\pi}\hat{E}(\alpha),\label{C5}\end{equation}
 where $\hat{K}(\alpha)$ and $\hat{E}(\alpha)$ are called complete
elliptic integrals of the first and second kinds, respectively.

If we take the derivative of $F_{1/2}$ with respect to $k\equiv\sqrt{\alpha}$
via Eq.~(\ref{C3}), we obtain\begin{equation}
\frac{{\partial F_{1/2}}}{\partial k}=-\frac{{F_{1/2}}}{k}+\frac{{F_{3/2}}}{k},\label{C6}\end{equation}
 or using Eq.~(\ref{C4})\begin{equation}
\frac{{\partial\hat{K}}}{\partial k}=-\frac{{\hat{K}}}{k}+\frac{{\pi}}{2}\frac{{F_{3/2}}}{k}.\label{C7}\end{equation}
 However, Ref.~\cite{arfken(01)} shows that \begin{equation}
\frac{{\partial\hat{K}}}{\partial k}=\frac{{\hat{E}}}{k\left(1-k^{2}\right)}-\frac{{\hat{K}}}{k}.\label{C8}\end{equation}
 Thus, by comparing Eq.~(\ref{C7}) and Eq.~(\ref{C8}) we find
the representation \begin{equation}
F_{3/2}=\frac{{2}}{\pi}\frac{{\hat{E}}}{1-k^{2}}=\frac{{2}}{\pi}\frac{{\hat{E}}}{1-\alpha}.\label{C9}\end{equation}

Further, we can also find the representation for $F_{5/2}$. First,
taking the derivative of $F_{3/2}$ with respect to $k\equiv\sqrt{\alpha}$
via Eq.~(\ref{C3}) gives\begin{equation}
\frac{{\partial F_{3/2}}}{\partial k}=-\frac{{3F_{3/2}}}{k}+\frac{{3F_{5/2}}}{k}.\label{C10}\end{equation}
 Also, using Eq.~(\ref{C9}) together with Eqs.~(\ref{C3})-(\ref{C5}),
another expression for the same derivative is obtained solely in terms
of complete elliptic integrals\begin{equation}
\frac{{\partial F_{3/2}}}{\partial k}=\frac{{2}}{\pi}\frac{{\left(1+k^{2}\right)\hat{E}-\left(1-k^{2}\right)\hat{K}}}{k\left(1-k^{2}\right)^{2}}.\label{C11}\end{equation}
 Then, by Eqs.~(\ref{C9}), (\ref{C10}), and, (\ref{C11}) we find\begin{equation}
F_{5/2}=\frac{{2}}{3\pi}\left[\frac{{2\left(2-\alpha\right)\hat{E}}}{\left(1-\alpha\right)^{2}}-\frac{{\hat{K}}}{1-\alpha}\right].\label{C12}\end{equation}

Now, using Eqs.~(\ref{C4}), (\ref{C9}) and (\ref{C12}), we may
rewrite the non-zero $B_{a}$ regularization parameters, Eqs. (\ref{eq:13})-(\ref{eq:15})
in Section \ref{sec:MODE-SUM-DECOMPOSITION-AND} as \begin{equation}
B_{t}=\frac{{q^{2}}}{r_{\mathrm{o}}^{2}}\frac{{E\dot{{r}}\left[\hat{K}(\alpha)-2\hat{E}(\alpha)\right]}}{\pi\left(1+J^{2}/r_{\mathrm{o}}^{2}\right)^{3/2}},\label{C13}\end{equation}

\begin{equation}
B_{r}=\frac{{q^{2}}}{r_{\mathrm{o}}^{2}}\frac{{\left(\dot{r}^{2}-2E^{2}\right)\hat{K}(\alpha)+\left(\dot{r}^{2}+E^{2}\right)\hat{E}(\alpha)}}{\pi\left(1-2M/r_{\mathrm{o}}\right)\left(1+J^{2}/r_{\mathrm{o}}^{2}\right)^{3/2}},\label{C14}\end{equation}

\begin{equation}
B_{\phi}=\frac{{q^{2}}}{r_{\mathrm{o}}}\frac{{\dot{r}\left[\hat{K}(\alpha)-\hat{E}(\alpha)\right]}}{\pi\left(J/r_{\mathrm{o}}\right)\left(1+J^{2}/r_{\mathrm{o}}^{2}\right)^{1/2}},\label{C15}\end{equation}
 which are exactly the same to the results of Barack and Ori \cite{barack-ori(02)}.


\begin{thebibliography}{10}
\bibitem{detweiler-whiting(03)}S. Detweiler and B. F. Whiting, Phys. Rev. D \textbf{67}, 024025 (2003),
gr-qc/0202086.
\bibitem{dirac(38)}P. A. M. Dirac, Proc. R. Soc. (London) \textbf{A167}, 148 (1938).
\bibitem{dewitt-brehme(60)}B. S. DeWitt and R. W. Brehme, Ann. Phys. \textbf{9}, 220 (1960).
\bibitem{mino-sasaki-tanaka(97)}Y. Mino, M. Sasaki, and T. Tanaka, Phys. Rev. D \textbf{55}, 3457
(1997).
\bibitem{quinn-wald(97)}T. C. Quinn and R. M. Wald, Phys. Rev. D \textbf{56}, 3381 (1997).
\bibitem{quinn(00)}T. C. Quinn, Phys. Rev. D \textbf{62}, 064029 (2000).
\bibitem{detweiler-m-w(03)}S. Detweiler, E. Messaritaki, and B. F. Whiting, Phys. Rev. D \textbf{67},
104016 (2003).
\bibitem{barack-ori(00)}L. Barack and A. Ori, Phys. Rev. D \textbf{61}, 061502(R) (2000).
\bibitem{mino-nakano-sasaki(02)}Y. Mino, H. Nakano, and M. Sasaki, Prog. Theor. Phys. \textbf{108},
1039 (2002), gr-qc/0111074.
\bibitem{barack-ori(02)}L. Barack and A. Ori, Phys. Rev. D \textbf{66}, 084022 (2002), gr-qc/0204093.
\bibitem{barack-mino-n-o-s(02)}L. Barack, Y. Mino, H. Nakano, A. Ori, and M. Sasaki, Phys. Rev. Lett.
\textbf{88}, 091101 (2002).
\bibitem{thorne-hartle(85)}K. S. Thorne and J. B. Hartle, Phys. Rev. D \textbf{31}, 1815 (1985).
\bibitem{zhang(86)}X.-H. Zhang, Phys. Rev. D \textbf{34}, 991 (1986).
\bibitem{MTW(73)}C. W. Misner, K. S. Thorne, and J. A. Wheeler, \emph{Gravitation}
(Freeman, San Fransisco, 1973).
\bibitem{weinberg(72)}S. Weinberg, \emph{Gravitation and Cosmology} (Wiley, New York, 1972).
\bibitem{jackson(99)}J. D. Jackson, \emph{Classical Electrodynamics} (Wiley, New York,
1999), 3rd ed.
\bibitem{mathews-walker(70)}J. Mathews and R. L. Walker, \emph{Mathematical Methods of Physics}
(W. A. Benjamin, New York, 1970), 2nd ed.
\bibitem{arfken(01)}G. B. Arfken and H. J. Weber, \emph{Mathematical Methods for Physicists}
(Harcourt/Academic Press, San Diego, 2001), 5th ed. \end{thebibliography}
\end{document}